\documentstyle[fleqn,twoside,epsfig]{article} 
\def\draft#1{}
\def\eqn#1{eq.~(\ref{#1})}

\def\fig#1{fig.~{\ref{#1}}}

\def\pol{\varepsilon}
\def\del{\partial}
\def\Tr{{\rm Tr}}
\def\tr{{\rm tr}}

\newbox\charbox
\newbox\slabox
\def\s#1{{      
        \setbox\charbox=\hbox{$#1$}
        \setbox\slabox=\hbox{$/$}
        \dimen\charbox=\ht\slabox
        \advance\dimen\charbox by -\dp\slabox
        \advance\dimen\charbox by -\ht\charbox
        \advance\dimen\charbox by \dp\charbox
        \divide\dimen\charbox by 2
        \raise-\dimen\charbox\hbox to \wd\charbox{\hss/\hss}
        \llap{$#1$}
}}
\def\I{{\cal I}}
\def\hf{\textstyle{1\over2}}
\def\eps{\epsilon}
\def\e{\epsilon}
\def\Dsl{\s D}
\def\si{\sigma}
\def\rg{r_{\Gamma}}
\def\cg{c_{\Gamma}}
\def\Ord{{\cal O}}

\def\tree{{\rm tree}}
\def\oneloop{{\rm 1\! -\! loop}}
\def\SUSY{{\rm SUSY}}
\def\neqfour{{N=4}}
\def\neqone{{N=1}}
\def\qb{\bar q}
\def\ksl{\s k}

\def\jb{{\bar\jmath}}
\def\Split{\mathop{\rm Split}\nolimits}
\def\Fact{\mathop{\rm Fact}\nolimits}

\def\spa#1.#2{\left\langle#1\,#2\right\rangle}
\def\spb#1.#2{\left[#1\,#2\right]}
\def\lor#1.#2{\left(#1\,#2\right)}
\def\sand#1.#2.#3{%
  \left\langle\smash{#1}{\vphantom1}\right|{#2}%
  \left|\smash{#3}{\vphantom1}\right\rangle}
\def\sandp#1.#2.#3{%
  \left\langle\smash{#1}{\vphantom1}^{-}\right|{#2}%
  \left|\smash{#3}{\vphantom1}^{+}\right\rangle}
\def\sandpp#1.#2.#3{%
  \left\langle\smash{#1}{\vphantom1}^{+}\right|{#2}%
  \left|\smash{#3}{\vphantom1}^{+}\right\rangle}
\def\sandmm#1.#2.#3{%
  \left\langle\smash{#1}{\vphantom1}^{-}\right|{#2}%
  \left|\smash{#3}{\vphantom1}^{-}\right\rangle}
\def\sandpm#1.#2.#3{%
  \left\langle\smash{#1}{\vphantom1}^{+}\right|{#2}%
  \left|\smash{#3}{\vphantom1}^{-}\right\rangle}
\def\sandmp#1.#2.#3{%
  \left\langle\smash{#1}{\vphantom1}^{-}\right|{#2}%
  \left|\smash{#3}{\vphantom1}^{+}\right\rangle}

\newskip\humongous \humongous=0pt plus 1000pt minus 1000pt
\def\caja{\mathsurround=0pt}
\def\eqalign#1{\,\vcenter{\openup1\jot \caja
        \ialign{\strut \hfil$\displaystyle{##}$&$
        \displaystyle{{}##}$\hfil\crcr#1\crcr}}\,}
\newif\ifdtup

\newcounter{eqnumber}
\renewcommand{\theeqnumber}{\arabic{eqnumber}}
\def\equn{
\refstepcounter{eqnumber}
\eqno({\rm \theeqnumber})
}

\begin{document}

\noindent hep-ph/9602280 \hfill UCLA/96/TEP/5

\noindent Saclay-SPhT-T96/10 \hfill SLAC--PUB--7111

\vskip .5 cm 

\noindent
{\huge PROGRESS IN ONE-LOOP QCD}

\vskip .2 cm 
\noindent
{\huge COMPUTATIONS%
\footnote{To appear in Annual Reviews of Nuclear and Particle Science (1996).}
}
\bigskip

\noindent
{\large {\it Zvi Bern}}

\medskip
\noindent
Department of Physics,
UCLA, Los Angeles, CA 90095

\medskip
\noindent
{\large {\it Lance Dixon}}

\medskip
\noindent 
Stanford Linear Accelerator Center, Stanford University,\\ 
Stanford, CA 94309

\medskip
\noindent
{\large {\it David A. Kosower}}

\medskip
\noindent 
Service de Physique Th\'eorique, 
Centre d'Etudes de Saclay,\\
F-91191 Gif-sur-Yvette cedex, France

\bigskip
\noindent
{\sc key words}:\quad  perturbative QCD, strings, unitarity, factorization

\vskip -.2 cm 
\bigskip

\tableofcontents
\bigskip

\hrule
\bigskip
\begin{abstract}
We review progress in calculating one-loop scattering amplitudes
required for next-to-leading-order corrections to QCD processes.  The
underlying technical developments include the spinor helicity
formalism, color decompositions, supersymmetry, string theory,
factorization and unitarity.  We provide explicit examples
illustrating these techniques.
\end{abstract}

\bigskip
\hrule
\bigskip



\section{INTRODUCTION}
\label{IntroductionSection}


\subsection{\it Importance of Diagrammatic Calculations}

Gauge theories form the backbone of the Standard Model.  The
weak-coupling perturbative expansion of gauge theory scattering
amplitudes, carried out by means of Feynman diagrams, has led to
theoretical predictions in remarkable agreement with high-energy
collider data~\cite{PrecisionEWReview}.  This high-precision agreement
places strong bounds on new physics.  In the strong-interaction sector
of the Standard Model --- described by quantum chromodynamics --- the
precision is not as great as in the electroweak sector.  QCD is
asymptotically free, so the strong coupling constant $\alpha_s$
becomes weak at large momentum transfers, justifying a perturbative
expansion~\cite{AsymptoticFreedom}.  Physical quantities do depend on
nonperturbative, long-distance QCD, in the form of quantities such as
parton distribution and fragmentation functions, as well as on the
physics of hadronization.  In many processes at modern colliders,
however, the dominant theoretical uncertainties are due to an
incomplete knowledge of the perturbation series, rather than to our
relative ignorance of nonperturbative aspects of scattering processes.
The situation is exacerbated by the slow approach to asymptopia
($\alpha_s$ is of order $0.1$ at the 100~GeV scale), and 
by the presence of large logarithms of ratios of scales.

\begin{figure}
\begin{center}
\epsfig{file=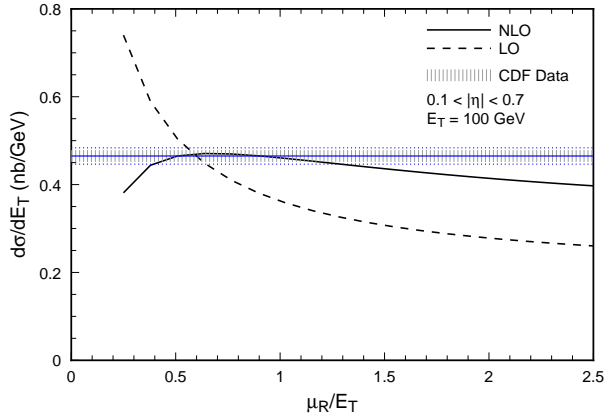,clip=,width=3.3in}
\end{center}
\vskip -.75 cm
\caption[]{
\label{ExptTheoryFigure}
The inclusive cross section for single-jet production in $p \bar p$
collisions at $\sqrt{s} = 1.8$~TeV and jet transverse energy
$E_T=100$~GeV (using MRS\,D${}_0'$ structure functions \cite{MRS}),
showing the sensitivity of the LO result to the choice of
renormalization scale, $\mu_R$, and the reduced sensitivity at NLO.  The CDF
data shown is extracted from ref.~\cite{CDFinclusiveJet}; the band
shows statistical errors only.}
\end{figure}

The leading-order (LO) term in the $\alpha_s$ expansion of a QCD
cross-section comes simply from squaring a tree-level scattering
amplitude.  Efficient techniques for computing QCD tree amplitudes
have been available for some time now~\cite{ManganoReview}, and the
results have provided a basic theoretical description of QCD processes
and thereby estimates of QCD backgrounds to new physics searches.
Unfortunately, higher order corrections, especially those enhanced by
logarithms, can be sizeable.  The ultraviolet logarithms manifest
themselves in the residual renormalization-scale dependence of a
finite-order prediction.  The renormalization scale $\mu_R$ is
introduced in order to define the coupling constant; renormalization
group invariance requires any physical quantity to be independent of
it.  However, when a perturbative expansion is truncated at a finite
order, residual $\mu_R$-dependence appears, because the cancellation
takes place across different orders in $\alpha_s$.  Calculations at
next-to-leading order (NLO) in $\alpha_s$ significantly reduce the
dependence on $\mu_R$ as compared to leading order.  As an example,
\fig{ExptTheoryFigure} shows the comparison of the LO and NLO
theoretical predictions to the experimental measurement of a point in
the single-jet inclusive distribution.  Note the good agreement
between NLO theory and experiment and the significant reduction of
theoretical uncertainties, compared to the LO calculation.

\def\yir{y_{\rm IR}}
Infrared logarithms 
arise because jet processes involve more than one scale, at the very least
a scale characterizing the jet size in addition to the hard scale of the
short-distance scattering, and because of the infrared divergences of
perturbative QCD.  These divergences transform the perturbation expansion
for such quantities from one in $\alpha_s$ alone to one in 
$\alpha_s \log^2 \yir$ and $\alpha_s\log \yir$ in addition to $\alpha_s$,
where $\yir$ is a jet `resolution' parameter.
All three must be small for the perturbation expansion to be reliable; but
the first two cannot be calculated in an LO calculation.  Only
in an NLO calculation are the corresponding terms determined
quantitatively, and only at this order can one establish the reliability of
the perturbative calculation.

Beyond the logarithmically-enhanced corrections, the ${\cal
O}(\alpha_s)$ corrections to most jet observables are larger than
non-perturbative power corrections and corrections due to quark
masses, and are thus the most important ones to calculate in order to
refine the precision of theoretical predictions.

%

Despite the need for higher-order QCD computations, at present no
quantities have been computed beyond next-to-next-to-lead\-ing order
(NNLO), and the only quantities that have been computed fully at NNLO
are totally inclusive quantities such as the total cross-section for
$e^+e^-$ annihilation into hadrons, and the QCD corrections to various
sum rules in deeply inelastic scattering \cite{eeNNLO,epNNLO}.  At
NLO, there are many complete calculations (in the form of computer
programs producing numerical results) for a variety of processes, but
at present results are still limited to where the basic process has
four external legs (counting electroweak vector bosons rather than
their decay products as external legs).  The following are examples of
calculations which are relevant for current experiments but have not
yet been performed or assembled:
\begin{enumerate}
\item NLO corrections to three-jet production at hadron
colliders.  These contributions would allow a measurement of $\alpha_s$
(via the three-jet to two-jet ratio) at the highest experimentally 
available momentum transfers, as well as next-to-leading-order studies
of jet structure.
\item NLO corrections to $W$ + multi-jet production at hadron
colliders.  These processes form a background to the $t$ quark signal
at Fermilab.
\item
NLO corrections to $e^+e^- \rightarrow 4$ jets.  At the $Z$ resonance,
this is the lowest-order process in which the quark and gluon color
charges can be measured independently.  It will also be useful for ruling
out the presence of light colored fermions (or scalars).
At LEP2 it is a background to threshold
production of $W$ pairs, when both $W$'s decay hadronically.
\item
NNLO corrections to $e^+e^-\rightarrow 3$ jets.  These corrections are 
the dominant uncertainty in a precision extraction of $\alpha_s$ from 
hadronic event shapes at the $Z$ \cite{ThreeJetUncertainty}.  
\end{enumerate}
In any of these processes, deviations of experimental results from the
theoretical predictions could indicate new physics.

\begin{figure}
\begin{center}
\epsfig{file=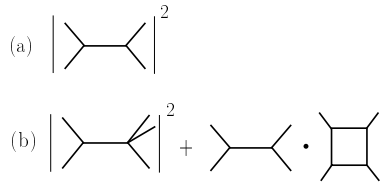, width=2in,clip=}
\end{center}
\vskip -.9 cm 
\caption[]{
\label{CrossSectionFigure}
In (a) the parton subprocesses required for the LO contribution to 
two-jet production at hadron colliders are shown schematically.  
In (b) the corresponding real and virtual NLO contributions are 
shown.}
\end{figure}

Why do these higher-order QCD corrections remain uncalculated?  NLO
corrections can be divided into real and virtual parts.  (See
\fig{CrossSectionFigure}.)  Real corrections arise from the emission
of one additional parton into the final state, and are straightforward
to compute from tree amplitudes with one more leg than the LO tree
amplitude.  Virtual corrections arise from the interference of the LO
tree amplitude with a one-loop amplitude.  Each contribution is
infrared divergent, but the divergences cancel in the sum, after
integrating the real contribution over ``unobserved'' partons in the
final state \cite{LeeNauenberg}, and factorizing initial state
singularities into the definition of parton distributions in an
incoming hadron \cite{FactorizationTheorem}.  The remaining finite
integrations are typically performed with a numerical program
\cite{JetPrograms}.

While the numerical evaluation of NLO corrections can be non-trivial,
the major analytical bottleneck is simply the availability of one-loop
amplitudes, which enter into the virtual corrections.  In particular,
one-loop amplitudes with more than four external legs (and all quarks
massless), which are required for the higher-order corrections listed
above, have only recently become available, thanks to the development
of new calculational techniques.  The purpose of this review is to
provide an introduction to some of these techniques, together with
worked-out examples.

%

Our emphasis will be on obtaining compact analytic results.  In
general, it is preferable to have such results for matrix elements,
even though they are ultimately inserted into numerical programs for
computing cross-sections.  Without compact results, numerical
instabilities can arise from the vanishing of spurious denominators in
the expression.  With analytic forms it is also easier to compare
independent calculations, to understand better how to organize
calculations, and even to obtain results for an arbitrary number of
external legs \cite{AllPlus,Mahlon,SusyFour,SusyOne}.


\subsection{\it Difficulty of Brute-Force Calculations}

Gauge theories have an elegant construction based on the principle of 
local gauge invariance.  The QCD Lagrangian for massless quarks $q$ is
$$ 
{\cal L}_{QCD} = -{1\over4}\Tr(F_{\mu\nu}^2) -i \bar{q}\Dsl q \,,
\equn\label{QCDL}
$$
where the covariant derivative 
$D_\mu = \partial_\mu - i g A_\mu/\sqrt{2}$ 
and field strength $F_{\mu\nu} = i\sqrt{2} [ D_\mu , D_\nu ]/g$ 
are given in terms of the matrix-valued gauge connection
$A_\mu = A_\mu^a T^a$.\footnote{  
The normalization $\Tr(T^aT^b) = \delta^{ab}$ of the
fundamental-representation generators $T^a$ accounts for the 
$\sqrt{2}$'s here; it serves to eliminate the $\sqrt{2}$'s from 
the partial amplitudes defined below.}
Since ${\cal L}_{QCD}$ depends on a single coupling constant $g$, 
all the interactions are dictated by gauge symmetry.  
Unfortunately, the Feynman diagram expansion does not respect this 
invariance, because the quantization procedure fixes the gauge
symmetry.  
%
%
Individual diagrams are not gauge invariant, 
and are often more complicated than the final sum over diagrams.  
The non-abelian gluon self-interactions coming from the cubic and 
quartic terms in \eqn{QCDL} have a complicated index structure and 
momentum-dependence.
So while it is straightforward in principle to compute both tree and
loop amplitudes by drawing all Feynman diagrams and evaluating them,
in practice this method becomes extremely
inefficient and cumbersome as the number of external legs grows.  
For five or more external legs there are a large number of kinematic
variables, which allow the construction of complicated expressions.
Indeed, intermediate expressions tend to be vastly more
complicated than the final results, when the latter are represented in
an appropriate way.

%
\begin{figure}
\begin{center}
\epsfig{file=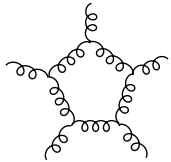, width=1.in,clip=}
\end{center}
\vskip -.7 cm 
\caption[]{
\label{FiveGluonFigure}
The five-gluon pentagon diagram.}
\end{figure}

As an example consider the five-gluon pentagon diagram, depicted
in \fig{FiveGluonFigure}, which would be encountered in a brute-force
computation of NLO corrections to three-jet production at 
a hadron collider.
Each of the five non-abelian three-point vertices in the diagram is 
given by
$$
V^{abc}_{\mu\nu\rho}(k,p,q) = f^{abc} \Bigl(
   \eta_{\nu\rho}(p-q)_\mu
+ \eta_{\rho\mu} (q-k)_\nu + \eta_{\mu\nu} (k-p)_\rho \Bigr) \,,
\equn\label{FeynmanVertex}
$$
where $f^{abc}$ are the $SU(3)$ structure constants, $k$, $p$ and $q$
the momenta, and $\eta_{\mu\nu}$ the Minkowski metric.  As the
non-abelian vertex has six terms, a rough estimate of the number of
terms is about $6^5$.  Each term is associated with a loop integral
which evaluates to an expression on the order of a page in length.
This means that one is faced with about $10^4$ pages of algebra for
this single diagram.  As bad as this brute-force approach might seem,
the situation is actually worse, because of the structure of the
results.  After evaluating the integrals and summing over a few
hundred more diagrams one obtains expressions of the form $\sum_i {N_i
\over D_i}$, where the factors $N_i$ are polynomials in the gluon
polarization vectors and external momenta, and the $D_i$ (polynomials
in the external invariants) are produced when the loop integrals are
reduced to a standard set of functions.  In general the $D_i$ contain
spurious kinematic singularities which cancel only after combining many
terms over a common denominator; this causes an explosion of
terms in the numerator.

In contrast to the complexity of intermediate expressions, the final
results can be strikingly simple.  For example, the five gluon
amplitudes which we shall describe in
section~\ref{SusyDecompositionSubsection} are remarkably compact.


\subsection{\it Non-traditional Approaches}
\label{NonTraditionalSubsection}

Substantial progress has been made in the past decade in improving the
calculation of tree-level amplitudes.  Four ideas which have played an
important role are the spinor helicity method for gluon polarization
vectors \cite{SpinorHelicity}, the color decomposition \cite{Color},
supersymmetry identities \cite{OldSWI,NewSWI}, and the Berends and
Giele recurrence relations \cite{Recursive}.  Although these ideas
form a basis for the one-loop techniques described here, they have
been extensively reviewed in ref.~\cite{ManganoReview,TasiLance}, and
we permit ourselves only a brief review below.

As illustrated by the pentagon example above, one-loop computations 
are significantly more complicated than tree computations, 
so further techniques are useful for preventing an explosion in algebra.  
The additional ideas which we shall discuss in this review 
involve string theory, supersymmetry, unitarity, and factorization.  
String theory, for example, suggests better gauge choices, 
a supersymmetric decomposition of amplitudes, and an improved
disentanglement of color and kinematics.  Approaches based on
unitarity and factorization make use of the analytic properties of
amplitudes to build further amplitudes using known ones. Since these
approaches use gauge-invariant quantities as the basic building blocks
of new amplitudes they tend to be extremely efficient. Although we
shall not discuss recursion relations here, Mahlon has made
considerable progress in applying these to 
one-loop amplitudes~\cite{Mahlon}.  
For simplicity, we demonstrate the methods for amplitudes
where all external particles are gluons, even though most of the
techniques (or analogs of them) can be applied to amplitudes 
with external fermions as well. 

To date, these techniques have allowed for the computation of all 
one-loop five-parton helicity 
amplitudes \cite{FiveGluon,Zoltanqqqqg,Fermion}, 
as well several infinite sequences of one-loop amplitudes
\cite{AllPlus,Mahlon,SusyFour,SusyOne}.  The five-parton amplitudes 
are currently being incorporated into numerical programs for NLO 
three-jet production at hadron colliders, the first item on the list in
Section 1.1 \cite{ZoltanProgram}.
Thus the analytical bottleneck to NLO corrections is yielding
to the new techniques described in this review.



\section{PRIMITIVE AMPLITUDES}
\label{ReviewSection}

In this section, we briefly review the use of color and helicity
information to decompose amplitudes into `primitive amplitudes'.
These building blocks have a much simpler analytic structure than the
full amplitudes, a fact which will be exploited in subsequent
sections.  We also review the application of supersymmetric Ward
identities to QCD.


\subsection{\it Color Decomposition}
\label{ColorSubsection}

Color decompositions have a long history, dating back to Chan-Paton
factors in early formulations of string theory \cite{Color}.
They are also related to the ``double-line'' formalism introduced by 
`t Hooft in the large-$N_c$ (number of colors) approach to
QCD~\cite{DoubleLine}, although here we will not make any large-$N_c$
or ``leading-color'' approximations.
The basic idea is to use group theory to break up an amplitude 
into gauge-invariant pieces which are composed of Feynman diagrams 
with a fixed cyclic ordering of external legs.
These pieces are simpler because poles and cuts can only appear in 
kinematic invariants made out of cyclicly adjacent sums of momenta,
of the form $(k_i+k_{i+1}+\cdots+ k_j)^2$.
At the four-point level this is not so important, because only one of
the three Mandelstam variables $s$,$t$,$u$ is thereby excluded;
but as the number of external legs grows, the total number of
invariants grows much faster than the number of cyclicly adjacent
ones.
The following brief review focuses on results needed later, 
rather than derivations.
A more complete discussion can be found in 
refs.~\cite{Color,ManganoReview,BKLoopColor,SusyFour,Fermion,TasiLance}.

We first generalize the gauge group of QCD to $SU(N_c)$, with the
quarks transforming in the fundamental representation.
The simplest way to implement the color decomposition in field theory
is by rewriting the group structure constants appearing in Feynman 
diagrams in terms of fundamental representation matrices
$$
f^{abc} = -{i\over\sqrt2} \Bigl( \Tr\bigl( T^a T^b T^c \bigr)
                           - \Tr\bigl( T^b T^a T^c \bigr) \Bigr) \,.
\equn
\label{FundConvert}
$$
After making this substitution in a generic Feynman diagram  
we obtain a large number of traces, many sharing 
$T^a$'s with contracted indices, of the form 
$\Tr\bigl(\ldots T^a\ldots\bigr)\,\Tr\bigl(\ldots T^a\ldots\bigr)
\,\ldots\,\Tr\bigl(\ldots)$.
If external quarks are present, then in addition to the traces there
will be some strings of $T^a$'s terminated by fundamental indices.
To reduce the number of traces and strings to a minimum, we rearrange
the contracted $T^a$'s, using
$$
  (T^a)_{i_1}^{~\bar j_1} \, (T^a)_{i_2}^{~\bar j_2}\ =\
  \delta_{i_1}^{~\bar j_2} \delta_{i_2}^{~\bar j_1}
  - {1\over N_c} \, \delta_{i_1}^{~\bar j_1} \delta_{i_2}^{~\bar j_2}\,,
\equn\label{ColorFierz}
$$
where the sum over $a$ is implicit.  (If all lines are in the adjoint
representation the second term drops out by a $U(1)$ decoupling
identity \cite{ManganoReview,Color}, which follows from the lack of a
`photon' self-coupling.)  A {\it partial amplitude} is the coefficient
of a given color trace in the resulting color decomposition of the
amplitude.

For example, in the $n$-gluon tree amplitude, application of 
\eqn{ColorFierz} reduces all color factors to single traces.  
Thus its decomposition is
$$
{\cal A}^\tree_n = g^{n-2} \sum_{\sigma\in S_n/Z_n} 
\Tr(\si(1)\ldots\si(n)) A^\tree_n(\si(1),\ldots,\si(n)) \,, 
\equn\label{TreeAmplitude}
$$
where $A_n^\tree$ are the partial amplitudes, 
$\Tr(1\ldots n) \equiv \Tr(T^{a_1}\ldots T^{a_n})$,
with $a_i$ the color index of the $i$-th external
gluon, and $S_n/Z_n$ is the set of non-cyclic
permutations of $\{1,2,\ldots,n\}$, corresponding to the set of
inequivalent traces.  
The labels on the gluon momenta $k_i$ and polarization vectors $\pol_i$,
implicit in \eqn{TreeAmplitude}, are also to be permuted by $\si$.
In the next subsection we will go over to a helicity basis, and
the label $i$ will be replaced by $i^{\lambda_i}$, with $\lambda_i$ the
(outgoing) gluon helicity. 
Similarly, tree amplitudes with
a pair of external quarks can be reduced to a sum over single
strings of matrices, $(T^{a_3}\cdots T^{a_n})_{i_2}^{~\jb_1}$,
and so on.  For a proof that individual partial amplitudes are
gauge invariant, see ref.~\cite{ManganoReview}.

At one loop, additional color structures are possible; 
in the $n$-gluon amplitude double traces appear as well as single
traces.  For example, the color decomposition of the one-loop
five-gluon amplitude is 
$$
\eqalign{
{\cal A}_{5}^\oneloop = & g^5 \mu_R^{2\e} \Biggl[
\sum_{\sigma \in S_5/Z_5}
N_c \Tr(\si(1)\ldots\si(5)) 
    A_{5;1} (\si(1),\ldots,\si(5)) \cr
& \hskip-16mm + \hskip-4mm
  \sum_{\sigma \in S_5/(S_2\times S_3)} \hskip-6mm
  \Tr(\si(1)\si(2)) \Tr(\si(3)\si(4)\si(5))
A_{5;3} (\si(1),\si(2);\si(3),\si(4),\si(5)) \Biggr] ; \cr}
\equn\label{LoopColor}
$$
as in~\eqn{TreeAmplitude} the permutation sums are over all 
inequivalent traces. 
For gauge group $U(N_c)$, the partial amplitudes $A_{5;2}$ multiplying
traces of the form $\Tr(1)\Tr(2345)$ would also have to be included,
but for $SU(N_c)$ the trace of a single generator vanishes. 
The decomposition of the $n$-gluon amplitude into single-trace
($A_{n;1}$) and double-trace ($A_{n;j>2}$) components is entirely
analogous.  Were one to consider the large-$N_c$ limit, the single-trace
terms would give rise to the leading contributions, and we will refer
to the corresponding partial amplitudes as leading-color partial amplitudes;
the double-trace terms have subleading-color partial amplitudes as 
coefficients.

\begin{figure}
\begin{center}
\epsfig{file=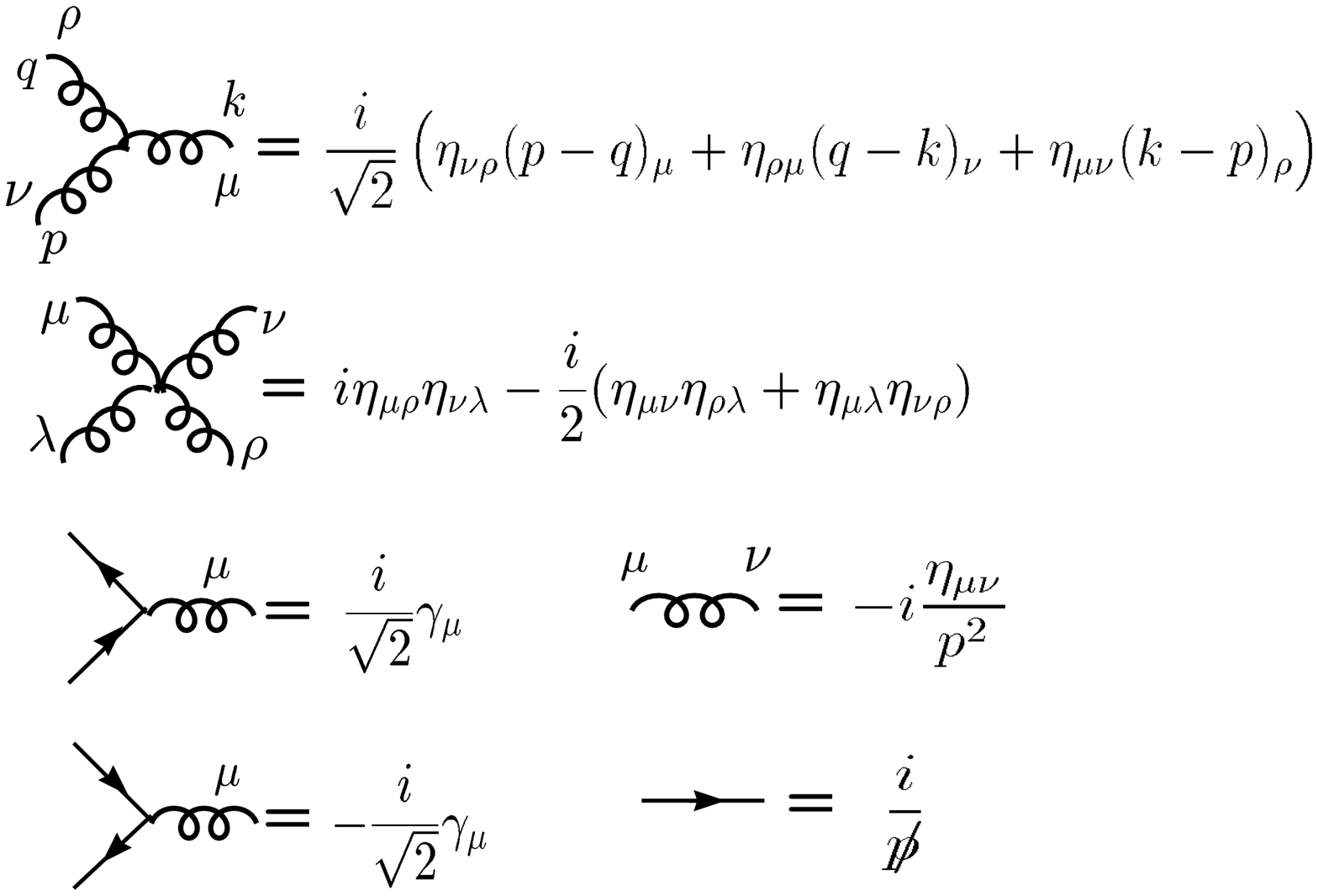,width=2.5in,clip=}
\end{center}
\vskip -.7 cm \caption[]{
\label{FeynmanColorFigure}
Color-ordered Feynman rules in Lorentz-Feynman gauge. Curly lines
represent gluons and lines with arrows fermions.}
\end{figure}

The rules for constructing leading-color partial amplitudes 
such as $A_n^\tree$ and $A_{n;1}$ are {\it color-ordered} Feynman 
rules, which are depicted in \fig{FeynmanColorFigure} for the standard
Lorentz-Feynman gauge.
These rules are obtained from
ordinary Feynman rules by restricting attention to a given ordering of
color matrices.  Applying \eqn{FundConvert} to \eqn{FeynmanVertex}
and extracting the coefficient of $\Tr(T^a T^b T^c)$ gives the 
color-ordered three-vertex in \fig{FeynmanColorFigure};
and similarly for the color-ordered four-vertex, 
the coefficient of $\Tr(T^a T^b T^c T^d)$. 
The only diagrams to be computed are those that can be drawn 
in a planar fashion with the external legs following the
ordering of the color trace under consideration.

%

\begin{figure}
\begin{center}
\epsfig{file=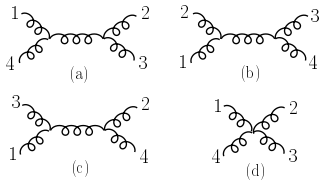,width=1.7in,clip=}
\end{center}
\vskip -.7 cm \caption[]{
\label{FourPointTreeFigure}
The four-point Feynman diagrams.  Color-ordered Feynman rules do not
include diagram (c) for $A_4(1,2,3,4)$.}
\end{figure}

The immediate advantage of rewriting Feynman rules in this way is that
fewer diagrams contribute to a given partial amplitude, and its
analytic structure is simpler.  As a simple example, with conventional
Feynman diagrams one would have a total of four conventional Feynman
diagrams, depicted in \fig{FourPointTreeFigure} for the four-point
tree amplitude.  With color-ordered Feynman rules one would compute
the partial amplitude $A_4(1,2,3,4)$ associated with the color trace
$\Tr(T^{a_1} T^{a_2} T^{a_3} T^{a_4})$, omitting
diagram~\ref{FourPointTreeFigure}c since the ordering of the legs do
not follow the ordering of the color trace.  Thus $A_4(1,2,3,4)$ has
no pole in $(k_1+k_3)^2$.  The other partial amplitudes can be
obtained by permuting the arguments of $A_4(1,2,3,4)$.  For the
five-gluon amplitude, there are 10 color-ordered diagrams as opposed
to 40 total.  Obviously the simplifications obtained using partial
amplitudes increase rapidly with the number of external legs.

%

At one loop, one also has to compute subleading-color partial
amplitudes, such as the double-trace coefficients $A_{5;3}$ in
\eqn{LoopColor}, which cannot be obtained directly from color-ordered
rules.  Fortunately there exist general formulas relating such
quantities to permutation sums of color-ordered objects
\cite{BKLoopColor,SusyFour,Fermion}.  For example, the gluon-loop
contribution to the four-gluon amplitude can be found from the
relation
$$
\eqalign{
A_{4;3}(1,2;3,4) &= 
  A_{4;1}(1,2,3,4) + A_{4;1}(1,3,2,4) + A_{4;1}(2,1,3,4)  \cr
& \hskip -1mm 
+ A_{4;1}(2,3,1,4) + A_{4;1}(3,1,2,4) + A_{4;1}(3,2,1,4)\,. \cr} 
\equn\label{FourPointPerms}
$$
Such formul\ae\ can be derived from string theory
\cite{Long,SusyFour}, although the most straightforward way to prove
them is using color flow diagrams in field theory \cite{Fermion}.  To
understand formula (\ref{FourPointPerms}) heuristically, it is useful
to focus on the box diagram.  Using ordinary Feynman rules and
expanding out the structure constants using eqs.~(\ref{FundConvert})
and (\ref{ColorFierz}) it is straightforward to check that the box
diagrams contribute to $A_{4;1}$ and $A_{4;3}$ in such a way
that~\eqn{FourPointPerms} is satisfied.  Roughly speaking, gauge
invariance then requires the remaining diagrams to tag along properly
with the box diagram.

Thus we can restrict our discussion henceforth to amplitudes with 
a fixed ordering of external legs, which we call 
{\it primitive amplitudes}.  
In the $n$-gluon cases discussed above, the set of primitive
amplitudes coincides with the leading-color partial amplitudes
$A_n^\tree$ and $A_{n;1}$, but this is not always the case.  
For example, one-loop amplitudes with external fermions have 
leading-color (as well as subleading-color) partial amplitudes 
that are sums of several primitive amplitudes \cite{Fermion}.

\subsection{\it Spinor Helicity Formalism}
\label{SpinorHelicitySubsection}

In explicit calculations, it is very convenient to adopt a 
helicity (circular polarization) basis for external gluons.
The spinor helicity formalism~\cite{SpinorHelicity} expresses
the positive- and negative-helicity polarization vectors
in terms of massless Weyl spinors $\vert k^{\pm} \rangle$,
$$
\pol^{+}_\mu (k;q) =  {\sandmm{q}.{\gamma_\mu}.k
      \over \sqrt2 \spa{q}.k}\, ,\hskip 1cm
\pol^{-}_\mu (k;q) =  {\sandpp{q}.{\gamma_\mu}.k
      \over \sqrt{2} \spb{k}.q} \, ,
\equn
\label{PolarizationVector}
$$
where $q$ is an arbitrary null `reference' momentum which drops out of
the final gauge-invariant amplitudes.  (Changing $q$ is equivalent to
performing a gauge transformation on the external legs.)
We use the compact notation
$$
\langle k_i^{-} \vert k_j^{+} \rangle \equiv \langle ij \rangle \, ,
\hskip 1 cm
\langle k_i^{+} \vert k_j^{-} \rangle \equiv [ij] \, .
\equn
$$
These spinor products are crossing-symmetric, antisymmetric in their
arguments, and satisfy
$$
\spa{i}.j \spb{j}.i = 2 k_i \cdot k_j \equiv s_{ij} \, .
\equn
$$
Helicity amplitudes can be given a manifestly crossing symmetric
representation, with the convention that a helicity label corresponds
to an outgoing particle; the helicity of an incoming particle is
reversed.  As we shall discuss in Section \ref{FactorizationSection}, in the
collinear limit where $k_i$ and $k_j$ become parallel, helicity
amplitudes have a square-root singular behavior, $\sim
{1\over\sqrt{s_{ij}}} \sim {1\over \spa{i}.{j}} \sim {1\over
\spb{i}.{j}}$, whose magnitude and phase are captured concisely by the
spinor products.  This helps explain why spinor products provide an
extremely compact representation of amplitudes.

In performing calculations, the Schouten identity is useful,
$$
\spa{i}.{j}\spa{k}.{l} 
= \spa{i}.{l}\spa{k}.{j} + \spa{i}.{k}\spa{j}.{l} \,.
\equn\label{SchoutenIdentity}
$$
A more complete discussion, including further identities and numerical
representations of the spinor products, can be found in
refs.~\cite{SpinorHelicity,ManganoReview,TasiLance}.

To maximize the benefit of the spinor helicity formalism for loop
amplitudes we must choose a compatible regularization scheme.  In
conventional dimensional regularization \cite{CollinsBook}, the
polarization vectors are $(4-2\eps)$-dimension\-al; this is
incompatible with the spinor helicity method, which assumes
four-dimensional polarizations.  To avoid this problem, we modify the
regularization scheme so all helicity states are four-dimensional and
only the loop momentum is continued to $(4-2\eps)$ dimensions.  This
is the four-dimensional-helicity (FDH) scheme \cite{Long}, which has
been shown to be equivalent \cite{KunsztFourPoint} to an appropriate
helicity formulation of Siegel's dimensional-reduction scheme
\cite{Siegel} at one-loop.  The conversion between schemes has been
given in ref.~\cite{KunsztFourPoint}, so there is no loss of
generality in choosing the FDH scheme.

\subsection{\it Parity and Charge Conjugation}
\label{PCSubsection}

The reader might worry that the color and helicity decompositions
will lead to a huge proliferation in the number of primitive or partial
amplitudes that have to be computed. 
In fact, this does not happen,
thanks to the group theory relations mentioned above, plus the 
discrete symmetries of parity and charge conjugation.
Parity simultaneously reverses all helicities in an amplitude;
\eqn{PolarizationVector} shows that it is implemented by
the exchange $\spa{i}.{j} \leftrightarrow \spb{j}.{i}$.
Charge conjugation is related to the antisymmetry of the color-ordered
rules; for pure-glue partial amplitudes it takes the form of a 
reflection identity,
$$
A_n^\tree(1,2,\ldots,n)\ =\ (-1)^n\ A_n^\tree(n,\ldots,2,1) \,.
\equn\label{reflectionid}
$$
For amplitudes with external quarks, it allows one to exchange a 
quark and anti-quark, or equivalently to flip the helicity on a 
quark line.

As an example,
with the use of parity and cyclic ($Z_5$) symmetry, we
can reduce the five-gluon amplitude at tree level to a combination
of just four independent partial amplitudes:
$$
\eqalign{  
& A_5^\tree(1^+,2^+,3^+,4^+,5^+)\,,\qquad
A_5^\tree(1^-,2^+,3^+,4^+,5^+)\,, \cr
& A_5^\tree(1^-,2^-,3^+,4^+,5^+)\,,\qquad 
A_5^\tree(1^-,2^+,3^-,4^+,5^+) \,. \cr}
\equn\label{treefiveg}
$$
Furthermore, the first two partial amplitudes here vanish (see below),
and there is a group theory ($U(1)$ decoupling) relation between the
last two \cite{ManganoReview,TasiLance}, so there is only one
independent nonvanishing object to calculate. At one loop there are
four independent objects --- \eqn{treefiveg} with $A_5^\tree$ replaced
by $A_{5;1}$ --- but only the last two contribute to the NLO
cross-section, due to the tree-level vanishings.  The explicit
expression for $A_{5;1}(1^-,2^-,3^+,4^+,5^+)$ is given in
section~\ref{SusyDecompositionSubsection}.

\subsection{\it Supersymmetry Identities}
\label{SWISubsection}

What does supersymmetry have to do with a non-supersymmetric theory
such as QCD?  The answer is that tree-level QCD is ``effectively''
supersymmetric~\cite{NewSWI}.  Consider an $n$-gluon tree amplitude.
It has no loops in it, so it has no fermion loops in it.  The fermions
in the theory might as well be in the adjoint representation, that is,
the theory might as well be a super Yang-Mills theory.  Pure-gluon
tree amplitudes in QCD are indeed identical to those in the
supersymmetric theory, and are thus related by supersymmetry to
amplitudes with fermions (the gluinos).  It is however more useful to
think of such relations as connecting {\it partial\/} amplitudes.
These relations are the so-called supersymmetric Ward identities
(SWI)~\cite{OldSWI,ManganoReview,TasiLance}.  They connect pure-gluon
partial amplitudes to partial amplitudes with a quark pair, because
after the color information has been stripped off, the latter are
identical to partial amplitudes with gluinos instead of quarks.  Using
the SWI saves computational labor~\cite{NewSWI}.

The SWI relate amplitudes with all external gluons, $g$, to amplitudes
where a pair of gluons is replaced by a pair of gluinos, $\Lambda$, or
a pair of complex scalars, $\phi$.  Specifically, the SWI that we
shall make use of in later sections are
$$
\eqalign{
A_n^\SUSY(g_1^\pm, g_2^+, \ldots , g_n^+) &= 0 \,, \cr
A_n^\SUSY(\Lambda_1^-, g_2^+, \ldots, g_{n-1}^+, \Lambda_n^+) &= 0 \,, \cr
A_n^\SUSY(\phi_1^-, g_2^+, \ldots, g_{n-1}^+, \phi_n^+) &= 0 \,, \cr
A_n^\SUSY(\Lambda_1^-, g_2^+, \ldots , g_j^-, \ldots, \Lambda_n^+)
  &= { \spa{j}.{n} \over \spa{j}.{1} }\,
  A_n^\SUSY( g_1^-, g_2^+ , \ldots, g_j^-, \ldots, g_n^+) \,, \cr
A_n^\SUSY(\phi_1^-, g_2^+, \ldots ,g_j^-, \ldots, \phi_n^+)
  &= { {\spa{j}.{n}}^2 \over {\spa{j}.{1}}^2 }
\, A_n^\SUSY( g_1^-, g_2^+ , \ldots, g_j^-, \ldots, g_n^+) \,, \cr}
\equn\label{SusyIdentities}
$$
where `$\ldots$' denotes positive-helicity gluons, and the helicity
assignments on $\phi$ refer to particle or antiparticle assignments
rather than genuine helicity.  At tree level, these identities hold
for all QCD partial amplitudes; for supersymmetric partial amplitudes,
they hold to all orders in perturbation theory.

At one loop QCD ``knows'' that it is not supersymmetric, but one can
still perform a supersymmetric decomposition of a QCD amplitude (see
section~\ref{SusyDecompositionSubsection}), for which the
supersymmetric components of the amplitude will
obey~\eqn{SusyIdentities}.  One may also use the identities to find
relations amongst non-supersymmetric contributions.  For example, in
$N=1$ super-Yang-Mills, one can use the first of the identities in
\eqn{SusyIdentities} to deduce that fermion- and gluon-loop
contributions are equal and opposite for $n$-gluon amplitudes with
maximal helicity violation. By considering an $N=2$ theory, with one
gluon, two gluinos and one (complex) scalar, one deduces that all three
types of loop contribution must be proportional to each other.  We
therefore obtain, for $SU(N_c)$ QCD with $n_s$ massless complex scalars and
$n_{\! f}$ massless Dirac fermions,
$$
A_{n;1}(g_1^\pm, g_2^+, \ldots , g_n^+) =
\Bigl(1 + {n_s \over N_c} - {n_{\! f} \over N_c} \Bigr)
 A_{n;1}^{\rm scalar}(g_1^\pm, g_2^+, \ldots , g_n^+) \,,
\equn
$$
where $A_{n;1}^{\rm scalar}$ is the contribution of a single scalar
and the factors of $1/N_c$ are the conversion factors between the
adjoint and fundamental representation loops.  (Note that
adjoint representation complex scalars have two states, 
but we have chosen the  normalization
that scalars in the fundamental representation 
--- $N_c\oplus\overline{N}_c$ --- have four states, 
the same as for their would-be superpartner fermions.)



\section{STRING-INSPIRED METHODS}
\label{StringInspiredSection}

String theory has provided a number of improvements in the calculation
of one-loop amplitudes.  Originally, we used it to derive a set of
diagrammatic computational rules for calculating gluon amplitudes
\cite{Long,StringBased}.  Such rules were used in the first
computation of the one-loop five-gluon amplitudes \cite{FiveGluon}.
The same methods also work well for gravity calculations
\cite{Gravity}.  One of the authors has reviewed the string-based
rules in ref.~\cite{TasiZvi} and we shall not do so here. Other
approaches to the string-based rules have been formulated
\cite{Lam,FirstQuantized}.  In particular, the first-quantized particle
world-line has led to a rather efficient computation of the two-loop
QED $\beta$-function \cite{TwoLoopQED} and of coefficients of
high-dimension operators in effective actions.


\subsection{\it String Organization}
\begin{figure}
\begin{center}
\epsfig{file=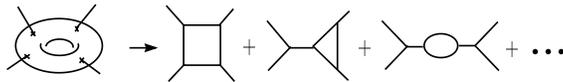,width=3.0in,clip=}
\end{center}
\vskip -.7 cm \caption[]{
\label{StringsFigure}
A single string diagram implicitly contains all field theory Feynman 
diagrams.}
\end{figure}

The basic motivation for the use of string theory follows from the compact
representation it provides for amplitudes:
at each loop order there is only a 
single closed string diagram.   As depicted in \fig{StringsFigure},
the string theory diagram contains within it all the Feynman diagrams,
including contributions of the entire tower of superheavy string
excitations.  The unwanted superheavy contributions are removed 
by taking the ``low-energy limit'' where all external momentum
invariants are much less than the string tension. 
This limit picks out different regions of integration in the string
diagram (see \fig{StringsFigure}), 
corresponding roughly to particle-like diagrams, but with
different, string-based, rules~\cite{StringBased}. 
 
%

Given knowledge of the string-based rules and organization, one may
also formulate a conventional field-theory framework which mimics them
\cite{Mapping} (at least for one-loop multiparton amplitudes), but
which can be applied more broadly (for example, to amplitudes with
external fermions).  At one loop, key ingredients of this
string-inspired framework are: use of a special gauge which is a
hybrid of Gervais-Neveu gauge~\cite{GN} and background-field
gauge~\cite{Background}; improved color decompositions; systematic
organization of the algebra; and a second-order formalism for fermions
\cite{Mapping,SecondOrder} which helps make supersymmetry relations
manifest.

\begin{figure}
\begin{center}
\epsfig{file=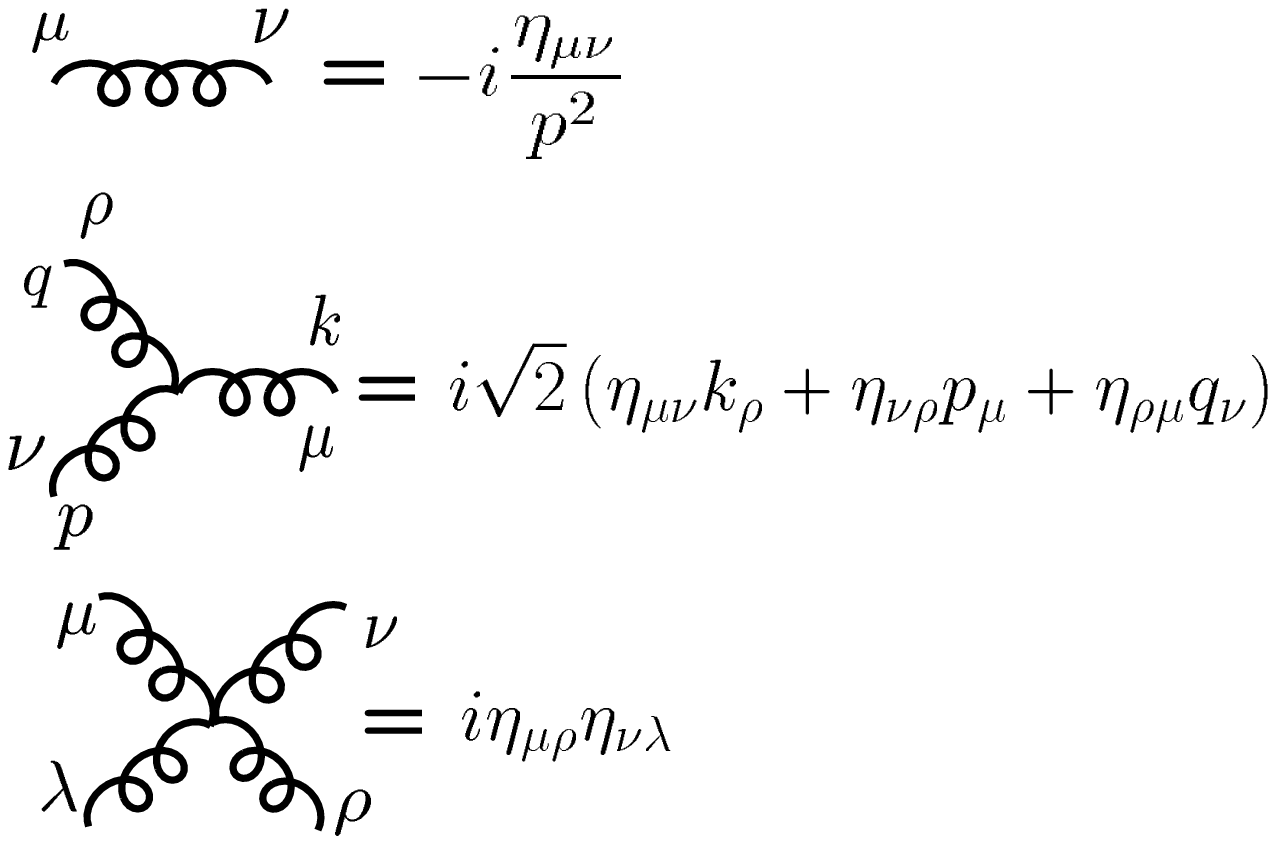,width=2.in,clip=}
\end{center}
\vskip -.7 cm \caption[]{
\label{GNRulesFigure}
The color-ordered Gervais-Neveu gauge three- and four-point vertices.}
\end{figure}

Gervais-Neveu gauge, originally derived from the low-energy 
limit of tree-level string amplitudes~\cite{GN}, has the following 
gauge-fixed action (ignoring ghosts):
$$
S^{\rm GN} = \int d^4 x \Bigl( -{1\over 4} \Tr[F^2_{\mu\nu}] -
{1\over 2} \Tr[(\partial \cdot A  - i g A^2/\sqrt{2})^2] \Bigr).
\equn\label{GNgauge}
$$
%
The color-ordered Feynman rules derived from this action are depicted
in \fig{GNRulesFigure}; comparing them to the color-ordered vertices
for the standard Lorentz-Feynman gauge (\fig{FeynmanColorFigure}),
we see that the three-point and four-point vertices have, respectively,
half and a third as many terms, showing why the Gervais-Neveu 
gauge is simpler for tree-level calculations.

%

\begin{figure}
\begin{center}
\epsfig{file=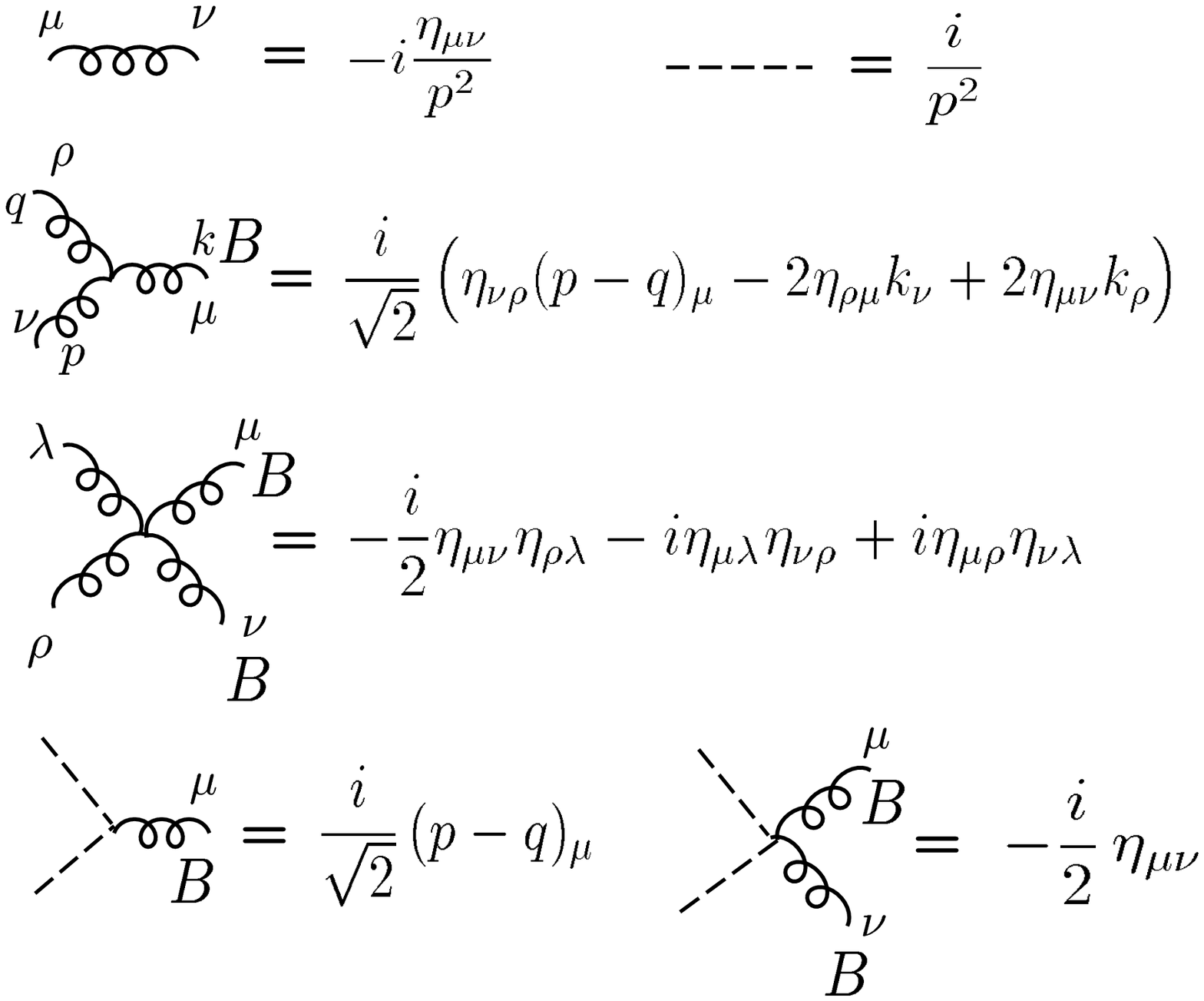,width=2.4in,clip=}
\end{center}
\vskip -.7 cm \caption[]{
\label{BackgroundRulesFigure}
The color-ordered background-field Feynman gauge three- and four-point 
vertices. Dashed lines represent either ghosts or scalars.}
\end{figure}

Given this understanding of the string reorganization of tree
amplitudes, one might guess that string theory would best be described
by the Gervais-Neveu gauge at one loop as well.  However, the gauge
most closely resembling the string organization of one-loop amplitudes
is a hybrid gauge involving both background-field and Gervais-Neveu
gauges~\cite{Mapping}.  To quantize in a background-field
gauge~\cite{Background} one splits the gauge field into a classical
background field and a fluctuating quantum field, $A_\mu\ =\ A_\mu^B +
A_\mu^Q$, and imposes the gauge condition $D_\mu^B A_\mu^Q = 0$, where
$D_\mu^B = \del_\mu - {i\over\sqrt{2}}gA_\mu^B$ is the
background-field covariant derivative, with $A_\mu^B$ evaluated in the
adjoint representation.  The Feynman-gauge
version of the gauge-fixed action is (again
ignoring ghosts),
$$
 S^{\rm Bkgd} = \int d^4 x \Bigl( 
  -{1\over 4} \Tr[F_{\mu\nu}^2] 
  -{1\over2}\Tr [(\partial\cdot A^Q 
        - i g[A_\mu^B,A_\mu^Q]/\sqrt{2})^2] \Bigr).
\equn\label{Backgroundgauge}
$$
The color-ordered background-field gauge vertices which arise
from expanding 
\eqn{Backgroundgauge} are depicted in \fig{BackgroundRulesFigure}.
Here we show only the vertices bilinear in the 
quantum field $A_\mu^Q$. These suffice for computing the one-loop
effective action $\Gamma[A^B]$, since $A_\mu^Q$ describes 
the gluon propagating around the loop while $A_\mu^B$ describes a 
gluon emerging from the loop.  

%

Any one-loop diagram can be split into a one-particle-irreducible
(1PI) part, or loop part, along with a set of tree diagrams sewn onto
the loop.  Now, $\Gamma[A^B]$ is invariant with respect to $A^B$ gauge
transformations~\cite{Background}.  Therefore we may use {\it any}
single gauge to compute the trees which are to be sewn onto the 1PI
parts of the diagrams.  Indeed, the string-motivated recipe is to use
background-field gauge only for the 1PI or loop vertices, and
Gervais-Neveu gauge for the remaining tree vertices~\cite{Mapping}.
This approach retains the above-noted advantages of Gervais-Neveu
gauge for tree computations, while avoiding the complicated ghost
interactions this nonlinear gauge would entail if it were used inside
the loop.  The advantage of the background-field gauge inside the loop
is that the loop momentum appears in only the first term in the
tri-linear gauge vertex in \fig{BackgroundRulesFigure}; the last two
terms contain only the external momentum $k$.  (In general, the most
complicated loop integrals to evaluate are those with the most
insertions of the loop momentum in the numerator.)  Furthermore, the
first term matches the scalar-scalar-gluon vertex, up to the
$\eta_{\nu\rho}$ factor.  Thus in background-field gauge the leading
loop-momentum behavior of one-particle-irreducible graphs with a gluon
in the loop is very similar to that of graphs with a scalar in the
loop.  Note also that the interactions of a scalar and of a ghost with
the background field are identical, up to the overall minus sign for a
ghost loop.  In the next subsection we elaborate further on these
relations.


\subsection{\it Supersymmetric Decomposition}
\label{SusyDecompositionSubsection}

String theory suggests a natural decomposition of QCD 
amplitudes into supersymmetric and non-supersymmetric parts.  
For example, for an $n$-gluon one-loop amplitude the contributions 
of a fermion and of a gluon circulating in the loop can be 
decomposed as
$$
\eqalign{
A_{n;1}^{\rm fermion}
&= -A_{n;1}^{\rm scalar} + A_{n;1}^\neqone\, , \cr
A_{n;1}^{\rm gluon}
&= A_{n;1}^{\rm scalar} - 4 A_{n;1}^\neqone + A_{n;1}^\neqfour\,.\cr}
\equn\label{TotalAmp}
$$
Here the ``scalar'' superscript denotes the contribution of
a complex scalar in the loop; the $N=1$ superscript refers 
to the contribution of a $N=1$ supersymmetric chiral multiplet,
consisting of a complex scalar and a Weyl fermion;
and the $N=4$ label refers to a vector supermultiplet,
consisting of three complex scalars, four Weyl fermions and 
a single gluon, all in the adjoint representation.  
(We have assumed the use of a supersymmetry preserving
regulator \cite{Siegel,Long,TasiZvi,KunsztFourPoint} in these
equations, and the vector-multiplet loop is defined to include the 
ghost loop.)

The two supersymmetric components of \eqn{TotalAmp} have important
cancellations in their leading loop-momentum behavior.
The simplest way to see this is via the scalar, fermion and 
gluon loop contributions to the background-field effective action,
$$
\eqalign{
\Gamma^{\rm scalar}[A] &= 
 \ln{\rm det}^{-1}_{[0]}\left( D^2 \right), \cr
\Gamma^{\rm fermion}[A] &= 
   \hf\ln{\rm det}^{1/2}_{[1/2]}
      \left( D^2- g
                \hf\sigma^{\mu\nu}F_{\mu\nu}/\sqrt{2} \right), \cr
\Gamma^{\rm gluon}[A] &= 
 \ln{\rm det}^{-1/2}_{[1]}
   \left( D^2- g\Sigma^{\mu\nu}F_{\mu\nu}/\sqrt{2}\right)
   + \ln{\rm det}^{}_{[0]}\left( D^2 \right)\,, \cr}
\equn\label{EffectiveAction}   
$$
where $\hf\sigma_{\mu\nu}$ and $\Sigma_{\mu\nu}$ are respectively the
spin-${1\over2}$ and spin-1 Lorentz generators, and where
${\rm det}{}_{[J]}$ is the one-loop determinant for a particle of
spin $J$ in the loop.
The fermionic contribution has been rewritten in second-order form
using 
$$
\eqalign{
   \ln{\rm det}^{1/2}_{[1/2]}\left( \Dsl \right) 
& = \hf\ln{\rm det}^{1/2}_{[1/2]}\left( \Dsl^2 \right) \,, \cr
 \Dsl^2 = \hf \{ \Dsl,\Dsl \} + \hf [ \Dsl,\Dsl ]   
& = D^2 - g \hf\sigma^{\mu\nu}F_{\mu\nu}/\sqrt{2} \,. \cr}
\equn\label{secondorder}
$$
In an $m$-point 1PI graph, the leading behavior of each contribution 
in \eqn{EffectiveAction} for large loop momentum $\ell$ is $\ell^m$.
The leading term always comes from the $D^2$ term in 
\eqn{EffectiveAction}, because $F_{\mu\nu}$ contains only the 
external momenta, not the loop momentum.  
Using $\Tr_{[0]}(1)=1,\ \Tr_{[1/2]}(1)=\Tr_{[1]}(1)=4$, we see that 
the $D^2$ term cancels between the scalar and
fermion loop, and between the fermion and gluon loop;
hence it cancels in any supersymmetric linear combination.
Subleading terms in supersymmetric combinations come from using 
one or more factors of $F$ in generating a graph; each $F$ costs one
power of $\ell$.   
Terms with a lone $F$ vanish, thanks to 
$\Tr\,\sigma_{\mu\nu} = \Tr\,\Sigma_{\mu\nu} = 0$.
This reduces the leading power in an $m$-point 1PI graph from $\ell^m$ 
down to $\ell^{m-2}$.  This argument can be extended to any amplitude
in a supersymmetric gauge theory \cite{SusyOne} and is related to the
improved ultraviolet behavior of supersymmetric amplitudes.
For the amplitude $A_{n;1}^\neqfour$, 
a comparison of the traces of products of two and three 
$\sigma_{\mu\nu}$'s and $\Sigma_{\mu\nu}$'s shows that  
further cancellations reduce the leading power behavior 
all the way down to $\ell^{m-4}$.  This result can
also be derived by superspace techniques~\cite{SuperSpace}.
In a gauge other than Feynman background-field gauge, the cancellations
involving the gluon loop would no longer happen diagram by
diagram. 

We illustrate the supersymmetric decomposition with the five-gluon 
amplitude, 
$A_{5;1}(1^-,2^-,3^+,4^+,5^+)$, 
whose components (\ref{TotalAmp}) are\,\cite{FiveGluon} 
$$
\eqalign{
  A^\neqfour &=  
     \cg \, A^\tree \sum_{j=1}^5 \Biggl[ 
     -{1\over\e^2}\! \left(-s_{j,j\!+\!1}\right)^{\!-\e}
  + \ln\left(\textstyle{-s_{j,j\!+\!1}\over -s_{j\!+\!1,j\!+\!2}}\right)
    \ln\left(\textstyle{-s_{j\!+\!2,j\!-\!2}\over -s_{j\!-\!2,j\!-\!1}}\right)
     \!+\! \textstyle{{\pi^2}\over6} \Biggr] \,,
\cr}
$$
$$
\eqalign{
  A^\neqone &=
     \cg \, A^\tree \Biggl[ {1\over\e} 
    -{1\over2}\left[ \ln(-s_{23})+\ln(-s_{51}) \right] + 2 \Biggr] 
\cr
&\hskip 10mm + {i\cg\over2}
   {{\spa1.2}^2 \bigl(\spa2.3\spb3.4\spa4.1+\spa2.4\spb4.5\spa5.1\bigr)
     \over \spa2.3\spa3.4\spa4.5\spa5.1}
     {\ln\left( {-s_{23}\over -s_{51}} \right) \over s_{51}-s_{23}} \,,
\cr}
$$
$$
\eqalign{
  A^{\rm scalar} &= 
   {1\over3} A^\neqone + {2\over9} \cg \, A^\tree - {i\cg\over3} 
\cr
&\hskip-12mm \times\!\Biggl[
    { \spb3.4\!\spa4.1\!\spa2.4\!\spb4.5\!
       \bigl(\! \spa2.3\!\spb3.4\!\spa4.1\!+\!\spa2.4\!\spb4.5\!\spa5.1\!\bigr)
          \over\spa3.4\spa4.5 } 
  { \ln\!\left(\!{-s_{23}\over -s_{51}}\!\right)
 \!-\!{1\over2}\!\left(\!{s_{23}\over s_{51}}\!-\!{s_{51}\over s_{23}}\!\right)
               \over (s_{51}-s_{23})^3 } 
\cr
& \hskip-10mm
   + {\spa3.5{\spb3.5}^3\over\spb1.2\!\spb2.3\!\spa3.4\!\spa4.5\!\spb5.1}
   \!-\! {\spa1.2{\spb3.5}^2\over\spb2.3\!\spa3.4\!\spa4.5\!\spb5.1} 
   \!-\! {1\over2}{\spa1.2\!\spb3.4\!\spa4.1\!\spa2.4\!\spb4.5\over
                  s_{23}\spa3.4\spa4.5 s_{51}} \Biggr] \,,\cr}
\equn\label{gggggmmppploop}
$$
where 
$$
 A^\tree \equiv A_5^\tree(1^-,2^-,3^+,4^+,5^+) 
  = i {\spa1.2^4\over \spa1.2 \spa2.3 \spa3.4 \spa4.5 \spa5.1} \,,
\equn\label{gggggmmppptree}
$$
$$
  \cg\ \equiv\ {\rg\over (4\pi)^{2-\e}}\ \equiv\  
   {\Gamma(1+\e)\Gamma^2(1-\e)\over(4\pi)^{2-\e}\Gamma(1-2\e)}\ .
\equn\label{cgammadef}
$$
These amplitudes contain both infrared and ultraviolet divergences,
which have been regulated dimensionally with $D=4-2\e$, retaining
terms through ${\cal O}(\e^0)$.  We see that the three components have
quite different analytic structure, indicating that the rearrangement
is a natural one.  The $N=4$ supersymmetric component is the simplest,
followed by the $N=1$ component.  The non-supersymmetric scalar
component is the most complicated, yet it is still simpler than the
amplitude with a gluon circulating in the loop, because it does not
mix all three components together.  The amplitudes for the one other
helicity configuration needed for NLO corrections,
$A_{5;1}(1^-,2^+,3^-,4^+,5^+)$, are a bit more complicated
\cite{FiveGluon}.

The surprising simplicity of $N=4$ supersymmetric loop amplitudes was
first observed by Green, Schwarz and Brink in their calculation of the
four-gluon amplitude as the low-energy limit of a superstring
amplitude~\cite{GSB}.  The supersymmetric decomposition can also
reveal structure in electroweak amplitudes that would otherwise remain
hidden~\cite{WeakInt}.  As we shall discuss in
section~\ref{Unitarity}, the cancellation of leading powers of loop
momentum for supersymmetric multiplets is extremely useful for
constructing such amplitudes via unitarity~\cite{SusyFour,SusyOne}.


\section{UNITARITY}
\label{Unitarity}

\def\hyph{\hbox{-}}
\def\scut{s\rm\hyph cut}
Unitarity has been a useful tool in quantum field theory since its
inception.  The Cutkosky rules \cite{Cutting,PeskinSchroeder} allow
one to obtain the imaginary%
\footnote{By imaginary we mean the discontinuities across branch cuts.}
(absorptive) parts of one-loop amplitudes directly from products of tree
amplitudes.  This is generally much easier than a full diagrammatic
calculation because one can greatly simplify the tree amplitudes {\it
before} feeding them into the calculation of the cuts.  

Having obtained the imaginary parts, one traditionally uses
dispersion relations to reconstruct real (dispersive) parts, up to
additive rational function ambiguities.  Although the Cutkosky rules
are computationally simpler than Feynman rules, the additive ambiguity 
has hampered their use in obtaining complete amplitudes.  Here we show 
how this problem is alleviated by the supersymmetry decomposition of 
section~\ref{StringInspiredSection}, and by a complete knowledge of all 
functions that may enter into a 
calculation~\cite{Integrals,IntegralRecursion}.

\subsection{\it Cutkosky Rules}

%
\begin{figure}
\begin{center}
\epsfig{file=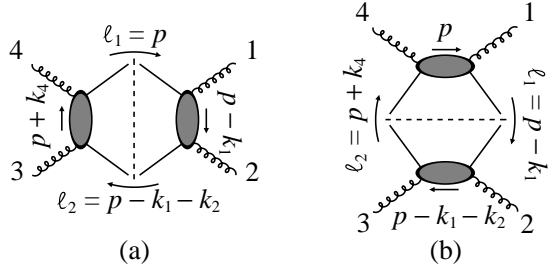,width=3.in,clip=}
\end{center}
\vskip -.7 cm \caption[]{
\label{FourPtCutFigure}
The $s$- and $t$-channel cuts of a one-loop four-gluon amplitude.
The cut lines can be gluons, fermions, or scalars.}
\end{figure}

Consider the $s$-channel cut of the four-point amplitude 
represented pictorially in \fig{FourPtCutFigure}a.  The Mandelstam
variables are as usual $s=(k_1+k_2)^2$ and $t=(k_2+k_3)^2$.
According to the Cutkosky rules, the $s$-channel cut (with $s>0$ and
$t<0$) of this amplitude is
$$
\eqalign{
-i\, {\rm Disc} \;
A_{4;1}(1,2,3,4)\Bigr|_{\scut} & = 
 \int {d^{4-2\eps} p \over (2\pi)^{4-2\eps}} \; 
 2\pi \delta^{\tiny(\!\tiny+\!)}(\ell_1^2)\, 2\pi\delta^{\tiny(\!\tiny+\!)}
 (\ell_2^2) \cr
& \hskip .3 cm \times
        A_4^\tree (-\ell_1, 1,2,\ell_2) \, 
        A_4^\tree (-\ell_2 ,3,4 ,\ell_1) \, ,  \cr}
\equn
\label{TreeProduct}
$$
where $\ell_1=p$ and $\ell_2 = p - k_1 - k_2$, 
$\delta^{\tiny(+)}$ is 
the positive-energy branch of the delta-function
and `Disc' means the discontinuity across the branch cut.
Color-ordering requires us to maintain the clockwise 
ordering of the legs in sewing the tree amplitudes.

Suppose the amplitude had the form $A_{4;1} = c \ln(-s) + \cdots 
= c (\ln|s| - i \pi) + \cdots$, 
where the coefficient $c$ is a rational function.
Then the phase space integral~(\ref{TreeProduct}) 
would generate the $ i \pi$ term but drop the
$\ln|s|$ term. Since we wish to obtain both types of terms, real and
imaginary, we replace the phase-space integral by the cut of an
unrestricted loop momentum integral \cite{SusyFour}; 
that is, we replace the $\delta$-functions with Feynman propagators, 
$$
\eqalign{
A_{4;1}&(1, 2, 3, 4)\Bigr|_{\scut}  =  \cr
& \hskip -3mm\left.
 \left[
\int\! {d^{4-2\eps}p\over (2\pi)^{4-2\eps}} \; 
  {i\over \ell_1^2 } \,
 A_4^\tree (-\ell_1, 1,2,\ell_2) \,{i\over \ell_2^2} \, 
        A_4^\tree (-\ell_2 ,3,4,\ell_1) \right]\right|_{\scut} \, .\cr}  
\equn
\label{TreeProductDef}
$$
While \eqn{TreeProduct} includes only imaginary parts,
\eqn{TreeProductDef} contains both real and imaginary parts.  As
indicated, \eqn{TreeProductDef} is valid only for those terms with an
$s$-channel branch cut; terms without an $s$-channel cut may not be
correct.  A very useful property of this formula is that one may
continue to use on-shell conditions for the cut intermediate legs
inside the tree amplitudes without affecting the result.  Only terms
containing no cut in this channel would change.  A similar equation
holds for the $t$-channel cut depicted in \fig{FourPtCutFigure}b.
Combining the two cuts into a single function, one obtains the full
amplitude, up to possible ambiguities in rational functions.

This procedure generalizes to an arbitrary number of external legs.
Isolate the cut in a single momentum channel by taking exactly one 
of the momentum invariants to be above threshold, and the rest of 
the cyclicly adjacent ones to be negative (space-like).  
To construct all terms with cuts in an amplitude, combine the 
contributions from the various channels into a single function with 
the correct cuts in all channels.  
Below we describe how to link the rational functions appearing in 
amplitudes to terms with cuts, so that complete
amplitudes can be obtained from Cutkosky rules.

\subsection{\it Cut Constructibility}

One-loop amplitudes satisfying a certain power-counting
criterion (for example supersymmetric amplitudes)
can be obtained directly from four-dimensional tree
amplitudes via the Cutkosky rules.  That is, when the criterion is
satisfied, one may fix all rational functions appearing in the
amplitudes directly from terms (through $\Ord(\eps^0)$) in the
amplitudes which contain cuts.  We refer to such amplitudes as
`cut-constructible'.  (Amplitudes not satisfying the criterion can 
still be obtained from cuts, but one must evaluate the cuts to 
higher order in $\eps$, which is more work.)
In the decomposition of $A_{5;1}$ given in \eqn{gggggmmppploop},
the $N=4$ and $N=1$ supersymmetric components are cut-constructible,
while the scalar component is not.  Correspondingly, rational
functions in the first two components (i.e., ${\pi^2\over6}$ and $2$)
are intimately linked to the logarithms, while the last three
rational terms in $A_{5;1}^{\rm scalar}$ are not so linked.

In a one-loop calculation one encounters integrals of the form 
$$
\eqalign{
{\cal I}_m[P(p^\mu)] & \equiv \int
{d^{4-2\e}p\over (2\pi)^{4-2\e}} \;
{P(p^\mu)\over p^2 (p-K_1)^2 (p-K_1-K_2)^2 \cdots (p+K_{m})^2}\, \cr
& \equiv {i(-1)^m\over(4\pi)^{2-\e}} I_m[P(p^\mu)]\,, }
\equn\label{GeneralMPoint}
$$
where $m$ is the number of propagators in the loop, $K_i$ are sums of
external momenta $k_i$, and $P(p^\mu)$ is the loop-momentum
polynomial.  A {\it cut-constructible} amplitude is one for which one
can arrange that all the $P(p^\mu)$ have degree at most $m-2$, except
for $m=2$ when $P$ should be at most linear.  Any amplitude satisfying
this power-counting criterion can be fully reconstructed from its cuts
(through $\Ord(\eps^0)$)~\cite{SusyOne}.  The basic idea behind the
proof is that only a restricted set of analytic functions appear in a
cut-constructible amplitude.  The standard Passarino-Veltman method
\cite{PV} reduces the generic tensor integral $I_m[P(p^\mu)]$ to a
linear combination of basic integrals with from 2 to $m$ external
legs.  (The kinematics of the lower-point integrals are obtained by
cancelling denominator factors in the original integral. In a
diagrammatic representation of the integrals, this corresponds to
pinching together adjacent external legs.)  A key feature of
Passarino-Veltman reduction is that integrals obeying the
power-counting criterion can be reduced entirely to scalar integrals
(integrals with $P=1$).  The proof of cut-constructibility is then
based on showing that the cuts provide sufficient information to fix
the coefficients of all the scalar integrals.  As we shall exemplify,
amplitudes not satisfying the power-counting criterion contain
additional rational functions, which spoil the argument.

\begin{figure}
\begin{center}
\epsfig{file=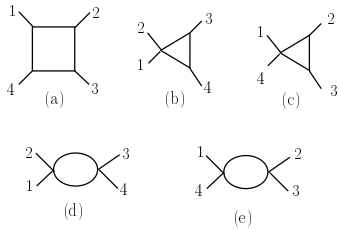,width=1.8in,clip=}
\end{center}
\vskip -.7 cm \caption[]{
\label{FourBasisFigure}
The independent scalar integrals that may appear in a massless
four-point calculation.}
\end{figure}

As an illustration, any cut-constructible massless four-point
amplitude must be given by a linear combination of the five scalar
integrals depicted in \fig{FourBasisFigure}.  (The triangle integral
with legs $3$ and $4$ pinched is equal to the integral with legs $1$
and $2$ pinched in \fig{FourBasisFigure}b and is therefore not
included in the figure; similarly, the one with $2$ and $3$ pinched is
equal to the one in \fig{FourBasisFigure}c.)  All these integrals
can be generated by Passarino-Veltman reduction of a box Feynman
diagram; the triangle and bubble integrals can also be generated from
other Feynman diagrams.  (Bubbles on external legs vanish in
dimensional regularization, and are therefore not included.) The
coefficients of the integrals are fixed by the cuts because each
integral contains logarithms unique to it: the box contains the
product $\ln(-s)\ln(-t)$, the triangles $\ln(-s)^2$ or $\ln(-t)^2$,
and the two bubbles contain $\ln(-s)$ or $\ln(-t)$. Consequently no
linear combination of these integrals with rational coefficients can
be formed which is cut-free.

The proof for an arbitrary number of external legs is similar,
although more complicated.  By systematically inspecting all scalar
integrals that enter into an $n$-point amplitude, one may
show that the cuts fix the coefficients of all integrals uniquely
\cite{SusyOne}.  One may also show that the errors induced by ignoring
the difference between using $D=4-2\eps$ and $D=4$ momenta on the cut
legs do not affect the cuts through $\Ord(\eps^0)$.
This observation is of considerable
practical use because $D=4$ tree amplitudes are simpler 
than those with legs in $D=4-2\eps$.

%

The proof breaks down for amplitudes that do not satisfy the
power-counting criterion.  For example, the scalar bubble with
momentum $K$,
$$
I_2[1](K) = {\rg\over\e(1-2\e)}(-K^2)^{-\e} 
          =  {1\over\e} + \ln(-K^2  ) + 2 + \Ord(\eps),
\equn\label{ScalarBubbleExample}
$$
obeys the criterion.  It contains a rational function, `2', but the latter is 
always accompanied by $\ln(-K^2)$.  
On the other hand, the linear combination  
$$
\Bigl( {K^{\mu}K^{\nu} \over 3} -{\eta^{\mu\nu} K^2 \over 12 } \Bigr)
I_2[1](K) -I_2[p^{\mu} p^{\nu}](K)
= -{ 1 \over 18} ( {K^{\mu}K^{\nu} }-\eta^{\mu\nu} K^2 ) + \Ord(\eps)
\equn\label{NoCutsExample}
$$
does not obey the criterion, because $I_2[p^{\mu} p^{\nu}](K)$ is
quadratic in the loop momentum.  The combination~(\ref{NoCutsExample})
is free of cuts through $\Ord(\eps^0)$; there is no logarithm attached to
it at this order.
The presence of such a combination within an amplitude cannot be detected 
using the $\Ord(\eps^0)$ cuts.  

In general, the power counting associated with a given amplitude
depends on the specific gauge choice and diagrammatic organization.
However, it suffices to find one organization of the diagrams satisfying
the power-counting criterion.  The string-inspired method discussed in
section~\ref{StringInspiredSection} provides such an organization; it
can satisfy the power-counting criterion even when the corresponding
diagrams in conventional Feynman gauge do not.  
An important class of cut-constructible amplitudes are those in supersymmetric 
gauge theory.  In section~\ref{SusyDecompositionSubsection} we showed
that for $n$-gluon amplitudes the leading two powers of loop momentum 
cancel in a supermultiplet contribution; the same result holds for
amplitudes with external fermions~\cite{SusyOne}.


\subsection{\it Supersymmetric Examples}
\label{SusyExamplesSubsection}

As a simple example, consider the contribution of an $N=4$
supersymmetry multiplet to a four-gluon amplitude.
This amplitude is an ordinary gauge-theory amplitude but with a 
particular matter content: 
one gluon, four gluinos and six real scalars all in the
adjoint representation.  As discussed in section~\ref{SWISubsection},
$A_{4;1}^{\SUSY}(1^\pm, 2^+, 3^+, 4^+) = 0$ so the first non-trivial
case to consider is $A_{4;1}^\neqfour(1^-, 2^-, 3^+, 4^+)$.

For the $s$-channel cut depicted in fig.~\ref{FourPtCutFigure},
only the gluon loop contributes; for fermion 
or scalar loops the supersymmetry identities in \eqn{SusyIdentities}
guarantee that at least one of the two tree amplitudes vanish.
The necessary tree amplitudes are the four-gluon amplitudes
$$
\eqalign{
& A_4^\tree(-\ell_1^+, 1^-, 2^-, \ell_2^+) = i { \spa1.2^4 \over
\spa{-\ell_1}.1 \spa1.2 \spa2.{\ell_2} \spa{\ell_2}.{-\!\ell_1}}\,, \cr
& A_4^\tree(-\ell_2^-, 3^+, 4^+, \ell_1^-) = i {
\spa{-\ell_1}.{\ell_2}^4 \over \spa{-\ell_2}.3 \spa3.4 \spa4.{\ell_1}
\spa{\ell_1}.{-\!\ell_2}} \,.\cr } 
\equn
$$
All other combinations of helicities of the intermediate lines cause
at least one of the tree amplitudes on either side of the cut to vanish. 
(The outgoing-particle helicity convention means that the helicity 
label for each intermediate line flips when crossing the cut.)
Cut-constructibility of supersymmetric amplitudes allows us to use the
four-dimensional tree amplitudes, so that
the cut in the $s$ channel, \eqn{TreeProductDef}, becomes
$$
\eqalign{
A_{4;1}^\neqfour(1^-, 2^-, 3^+, 4^+)\Bigr|_{\scut}& = 
  \int {d^{4-2\eps} p \over (2\pi)^{4-2\eps}} \; {i \over \ell_1^2} \;
{i \spa1.2^4 \over \spa{\ell_1}.1 \spa1.2 \spa2.{\ell_2} 
             \spa{\ell_2}.{\ell_1}} \cr
& \hskip 20 mm  \times \left.
{i\over \ell_2^2} \; {i\spa{\ell_1}.{\ell_2}^4
\over \spa{\ell_2}.3 \spa3.4 \spa4.{\ell_1} \spa{\ell_1}.{\ell_2}}
\right|_{s - \rm cut} \hskip -.6 cm \,, \cr}
\equn\label{SCutSusyA}
$$
where we have removed the minus signs from
inside the spinor products by cancelling constant phases.  To put this
integral into a form more reminiscent of integrals encountered in
Feynman diagram calculations we may rationalize the denominators
using, for example,
$$
 {1\over \spa2.{\ell_2}} = -{\spb2.{\ell_2} \over (p - k_1)^2} \, .
\equn\label{Rationalize}
$$
We use the on-shell conditions $\ell_1^2=0$ and 
$\ell_2^2=0$, which apply even though the loop integral is 
unrestricted, because of the $\scut$ restriction.
Performing such simplifications yields, 
$$
\eqalign{
A_{4;1}^\neqfour&(1^-, 2^-, 3^+, 4^+)\Bigr|_{\scut} \cr
& \hskip -.3 cm 
= - i A_4^\tree  \!\left.\left[ \int\!
   {d^{4-2\eps}p\over (2\pi)^{4-2\eps}} \;   
{ {\cal N} \over p^2 (p - k_1)^4 (p - k_1 - k_2)^2 (p + k_4)^4} 
\right]\right|_{\scut} \hskip - .6 cm \,, \cr}
\equn\label{SCutSusyB}
$$
where we have extracted a factor of the tree amplitude,
$$
A_{4}^\tree(1^-, 2^-, 3^+, 4^+) =
 i {\spa1.2^4 \over \spa1.2 \spa2.3 \spa3.4 \spa4.1} \,,
\equn\label{ggggmmpptree}
$$
from the amplitude.   The numerator of the integrand is
$$
\eqalign{
{\cal N} & = \spb{\ell_1}.1 \spa1.4 \spb4.{\ell_1} \spa{\ell_1}.{\ell_2}
             \spb{\ell_2}.3 \spa3.2 \spb2.{\ell_2} \spa{\ell_2}.{\ell_1} \cr
& = \tr_+ [\ell_1 1 4 \ell_1 \ell_2 3 2 \ell_2] \cr
& = - 4 \tr_+[4321]\,  \ell_1 \cdot k_1 \, \ell_1\cdot k_4
           = -st \, (p - k_1)^2 (p + k_4)^2 \,, \cr}
\equn
$$
where $\tr_+[\cdots] = {1\over 2} \tr[(1+\gamma_5) \cdots]$ 
and we used
$$
\ell_1^2 = 0 \, , 
\hskip 1 cm \s\ell_1 \s\ell_2 = \s\ell_1 (\s k_3 + \s k_4) \, , 
\hskip 1 cm \s\ell_2 \s\ell_1 = -(\s k_1 + \s k_2) \s\ell_1 \, . 
\equn
$$
The $\gamma_5$ term in the trace drops out because a four-point
amplitude has only three independent momenta to contract into the
totally anti-symmetric Levi-Civita tensor.

Thus in \eqn{SCutSusyB} the numerator neatly reduces the squared 
propagators to single propagators,
$$
i s t A_4^\tree \int
 {d^{4-2\eps}p\over (2\pi)^{4-2\eps}} \;
{1\over p^2 (p - k_1)^2 (p - k_1 - k_2)^2 (p + k_4)^2} 
\, \biggr|_{\scut}\,,  
\equn
$$
which is a scalar box integral.  Thus the $s$-cut contribution is
given by 
$$
A_{4;1}^\neqfour(1^-, 2^-, 3^+, 4^+)\Bigr|_{\scut} 
= {- s t \over (4\pi)^{2-\eps}}\, A_4^\tree  \, I_4(s,t)
 \Bigr|_{\scut} \,,
\equn
$$
where the massless scalar box integral is 
(see e.g. ref.~\cite{IntegralRecursion})
$$
I_4(s,t) = - {2 \rg \over st}  \biggl\{
 - {1\over\e^2}\! \Bigl[ (-s)^{-\e} 
                       + (-t)^{-\e} \Bigr]
  + {1\over 2} \ln^2\left({s\over t}\right) + {\pi^2\over 2} \biggr\}
+ \Ord(\eps) \,.
\equn\label{BoxIntegral}
$$ 

The evaluation of the $t$-channel cut depicted in
\fig{FourPtCutFigure} is similar, but a bit more involved since all
particles in the multiplet contribute.  However, after summing over
the contribution of all particles, with the help of the
SWI~(\ref{SusyIdentities}) and the Schouten
identity~(\ref{SchoutenIdentity}), the integral appearing in the
$t$-channel cut turns out to be the same as the one appearing in the
$s$-channel cut in \eqn{SCutSusyB}.

Combining the $s$ and $t$ channel results, the amplitude must be 
$$
A_{4;1}^\neqfour(1^-,2^-,3^+,4^+)
= {- s t \over(4\pi)^{2-\eps}}\, A_4^\tree  \, I_4(s,t) \, .
\equn\label{ggggmmppneqfour}
$$ 
The rational function proportional to $\pi^2$ contained in the box integral 
(\ref{BoxIntegral}) is fixed by the cuts since it appears in
association with the logarithms in this function.
Integrals having cuts in multiple channels, such as
$I_4(s,t)$, provide a strong consistency check: their coefficients can
be obtained via two or more separate cut calculations and the results
must agree.

\begin{figure}
\begin{center}
\epsfig{file=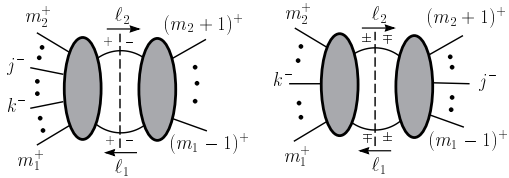,width=2.7in,clip=}
\end{center}
\vskip -.7 cm \caption[]{
\label{AllnFigure}
The relevant cuts for computing the MHV 
amplitudes for an arbitrary number of external legs.}
\end{figure}

\begin{figure}
\begin{center}
\epsfig{file=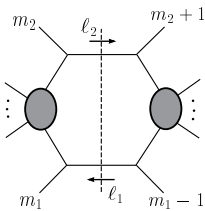,width=1.1in,clip=}
\end{center}
\vskip -.7 cm \caption[]{
\label{HexagonFigure}
All-$n$ MHV supersymmetric amplitudes can be evaluated by evaluating
hexagon integrals.}
\end{figure}

Following the same procedure one may evaluate the other nonvanishing
$N=4$ four-gluon amplitude, $A_{4;1}^\neqfour(1^-, 2^+, 3^-, 4^+)$,
where the negative helicities are non-adjacent.  Surprisingly, the
same basic calculation can be easily extended to an arbitrary number
of external legs for maximally helicity violating (MHV) amplitudes,
those with two negative-helicity gluons and the remaining of positive
helicity.  (A special case is $A_{5;1}^\neqfour(1^-,2^-,3^+,4^+,5^+)$,
given in~\eqn{gggggmmppploop}.)  The cuts fall into two categories,
depending on whether the external negative-helicity gluons are on the
same or on opposite sides of the cut, as depicted in \fig{AllnFigure}.
In either case the tree amplitudes on both sides of the cuts are given
by the Parke-Taylor formula \cite{ParkeTaylor,ManganoReview},
$$
\eqalign{
A^\tree(\ell_1^+,  m_1^+, & \ldots, k^-, \ldots, j^-, 
                      \ldots, m_2^+,\ell_2^+) \cr
& = i { \spa{k}.j^4 \over 
               \spa{\ell_1}.{m_1} \spa{m_1,}.{m_1\!+\!1} \cdots 
                     \spa{m_2\!-\!1,}.{m_2} \spa{m_2}.{\ell_2} 
                     \spa{\ell_2}.{\ell_1}} \,,\cr}
\equn\label{PTAmplitudes}
$$
where $j$ and $k$ are the two negative-helicity legs, or by formul\ae\
related to \eqn{PTAmplitudes} by the SWI~(\ref{SusyIdentities}).
The key to evaluating the cut integrals
for an arbitrary number of external legs is that only two denominator
factors in the tree amplitudes~(\ref{PTAmplitudes}) contain the loop
momentum (since $1/\spa{\ell_2}.{\ell_1} =\spb{\ell_2}.{\ell_1}/(k_{m_1}
+ \cdots+ k_{m_2})^2$).  Thus each tree contributes only two
propagators containing the loop momentum, so after including the two
cut propagators the hardest integral to be evaluated is the hexagon
integral depicted in \fig{HexagonFigure}.  These hexagon integrals can
be reduced to scalar box integrals in much the same way as for the
four-point case, allowing one to obtain the amplitudes for an
arbitrary number of external legs \cite{SusyFour}.

%
%

%

The analysis of $N=1$ supersymmetric MHV amplitudes is similar,
although more complicated \cite{SusyOne}.  Again the key to the
construction is that no more than six denominators contain loop
momentum, even for an arbitrary number of external legs.  One instance
of the general $N=1$ MHV result is provided by
$A_{5;1}^\neqone(1^-,2^-,3^+,4^+,5^+)$ in~\eqn{gggggmmppploop}.
Notice that only the $s_{23}$ and $s_{51}$ channels contain cuts.
This result (which is also true for the scalar component) is a simple
consequence of the supersymmetry identities~(\ref{SusyIdentities}).
The construction of amplitudes via cuts does not rely on
supersymmetry, but only on the power-counting criterion; however,
non-supersymmetric amplitudes generally do not satisfy the criterion.

\subsection{\it Non-supersymmetric Example}
\label{NonSusyExampleSubsection}

Amplitudes not satisfying the power-counting criterion require
an extension of this approach.  Consider the non-supersymmetric amplitude 
for four identical helicity gluons with a scalar in the loop, 
$$
A_{4;1}^{\rm scalar}(1^+, 2^+, 3^+, 4^+)
 =  -{i\over48\pi^2}
    {\spb1.2 \spb3.4 \over \spa1.2 \spa3.4 } + \Ord(\eps) \,,
\equn\label{FourPlusResult}
$$
first obtained from string-based techniques \cite{Long}.  
At first sight one might think that it is impossible to use unitarity 
to obtain this amplitude, since it contains no cuts.
The box Feynman diagram for this amplitude contains up to four powers
of loop momentum, so the power-counting criterion is not satisfied 
(in any gauge). 

However, in $D=4-2\eps$ all terms in a massless
amplitude necessarily have cuts \cite{TwoLoopUnitarity}: 
by dimensional analysis of~\eqn{GeneralMPoint}, all terms must be 
proportional to factors of $(-K^2)^{-\eps}$, where $K^2$ is some 
kinematic variable.  
In particular a massless four-point amplitude must be of the form
$$
\eqalign{ A_4^{D=4-2\eps} 
& = (-s)^{-\eps}\, f_1 + (-t)^{-\eps} \, f_2\cr 
& = (1-\eps \ln(-s)) f_1 + (1-\eps \ln(-t)) f_2 + \cdots \cr}
\equn\label{PolyCuts}
$$
where $f_1$ and $f_2$ are dimensionless functions of the kinematic
variables. This expression now contains cuts at $\Ord(\eps)$ even if
$f_1$ and $f_2$ are cut-free. Rational functions such as those in
\eqn{FourPlusResult} may therefore be obtained from the sum $f_1+f_2$,
fixed by the coefficients of the single logarithms at $\Ord(\eps)$.

\def\mub{\mu} 
Thus, to obtain the rational function contributions in
amplitudes which do not satisfy the power-counting criterion we must
perform a cut calculation valid to at least one higher order in
$\eps$.  We are not actually interested in the explicit values of the
$\Ord(\eps)$ terms; we only need to extract the sum $f_1 + f_2$.  To
implement a calculation valid to higher orders in $\eps$ we correct
for the fact that the loop momenta appearing in the tree amplitudes on
either side of the cut are in $(4-2\eps)$-dimensions instead of
four-dimensions.  The proper on-shell conditions on the cut legs are
$\ell_1^2 - \mub^2 = 0$ and $\ell_2^2 - \mub^2 = 0$, where $\ell_1$
and $\ell_2$ are left in four-dimensions and $\mub$ is the
$(-2\eps)$-dimensional part of the loop momentum.  We follow the
standard prescription that the $(-2\eps)$-dimensional subspace is
orthogonal to the four-dimensional one \cite{CollinsBook}.  For
practical purposes we may think of $\mub^2$ as a mass which gets
integrated over. (This decomposition of the loop momentum has also
been used by Mahlon \cite{Mahlon} in his recursive approach.)

For the amplitude in \eqn{FourPlusResult}, 
the tree amplitudes entering the two sides of the $s$-channel cut,
depicted in \fig{FourPtCutFigure}a, are easily computed from 
color-ordered Feynman diagrams; the one on the left side of the cut is 
$$
\eqalign{
A_4^\tree (-\ell_1, 1^+,2^+,\ell_2) & = 
i \mub^2 {\spb1.2 \over \spa1.2  ((\ell_1 - k_1)^2 - \mub^2)} \, , \cr}
\equn\label{MassiveScalarTree}
$$
where legs $\ell_1$ and $\ell_2$ represent the cut scalar lines.  The
one on the right side is obtained by relabeling legs.  The
amplitude~(\ref{MassiveScalarTree}) vanishes in $D=4$ ($\mub^2\to0$)
by the SWI~(\ref{SusyIdentities}).  Plugging these tree amplitudes
into \eqn{TreeProductDef}, we obtain the $s$ cut of the scalar loop
contribution,
$$
\eqalign{
A_{4;1}^{\rm scalar}&(1^+, 2^+, 3^+, 4^+)\Bigr|_{\scut} 
= 2\, {\spb1.2 \spb3.4 \over \spa1.2 \spa3.4 } \;
\I_4[\mub^4] \Bigr|_{\scut} \,, \cr}
\equn\label{ScalarSCut}
$$
where the factor of 2 is from the two states of a complex
scalar and
$$
\I_4[\mub^4] \equiv 
\int\! {d^{4} p \over (2\pi)^{4}}  {d^{-2\eps}\mub \over (2\pi)^{\!-2\eps}} \; 
        {\mub^4 \over (p^2 - \mub^2) ((p\! -\!k_1)^2 - \mub^2) 
                 \cdots ((p\! +\! k_4)^2 - \mub^2)} \,.
\equn\label{IntegralMuFour}
$$

The $t$-channel cut, depicted in \fig{FourPtCutFigure}b, is similar
and may be obtained via the relabeling $1\leftrightarrow 3$.
Using the identity
$$
{\spb3.2 \spb1.4 \over \spa3.2 \spa1.4 } = 
{\spb1.2 \spb3.4 \over \spa1.2 \spa3.4 } \, , 
\equn
$$
the $t$-cut is given simply by~\eqn{ScalarSCut}, with `$s$-cut' replaced by
`$t$-cut'.

Combining the two cuts we obtain an expression valid for both cuts,
$$
A_{4;1}^{\rm scalar}(1^+, 2^+, 3^+, 4^+)
= 2
{\spb1.2 \spb3.4 \over \spa1.2 \spa3.4 } \, \I_4[\mub^4] \,.
\equn\label{FinalScalar}
$$
Although we only calculated the cuts, we did so to all orders in $\e$; 
therefore by \eqn{PolyCuts} we know the complete loop amplitude.
To obtain the amplitude through $\Ord(\eps^0)$ we 
need only evaluate the leading $\Ord(\eps^0)$ contribution to the integral
$\I_4[\mub^4]$.

A good way to evaluate the leading term is to first
integrate out the angles in the $(-2\eps)$-dimensional subspace.
Using the fact that the integrand is a function only of $\mub^2$, we 
have 
$$
\int {d^{-2\eps} \mub \over (2\pi)^{-2\eps}}
\rightarrow  - (4\pi)^\eps {\eps \over \Gamma(1-\eps)} 
\int_0^\infty d \mub^2\; (\mub^2)^{-1-\eps} \,.
\equn
$$
The overall $\eps$ from the measure must be compensated by a $1/\e$
ultraviolet pole in the remaining integration.  
As usual a leading ultraviolet divergence may be extracted 
conveniently by setting all external momenta to zero and 
inserting a mass parameter $\lambda$ to replace the momenta.  
Thus we may evaluate the integral (\ref{IntegralMuFour}) as
$$
\eqalign{
\I_4[\mub^4] \rightarrow &\!
\int  {d^{4} p \over (2\pi)^{4}}  {d^{-2\eps} \mub \over (2\pi)^{-2\eps}} \; 
        {\mub^4 \over (p^2 - \mub^2 -\lambda^2)^4} \cr
& = - {i\eps \over (4 \pi)^{2-\eps}} {1\over \Gamma(1-\eps)}
    \int_0^\infty dp^2 
    \int_0^\infty d\mub^2 \; {p^2 \,(\mub^2)^{1-\eps} 
                    \over (p^2 + \mub^2 + \lambda^2)^4}\cr
& =  - {i\eps \over (4 \pi)^{2-\eps}} \Bigl({1\over 6\eps} + \Ord(1)
     \Bigr)\,, \cr}
\equn
$$
where we have used standard formulas \cite{PeskinSchroeder} for the
angular integrals and then integrated the radial dimension.  Plugging
the leading-in-$\e$ result into \eqn{FinalScalar}, we obtain the
correct result for the amplitude (\ref{FourPlusResult}).

Although this method can in principle be applied to any massless
one-loop amplitude to obtain complete amplitudes, it is generally
advantageous to first decompose amplitudes into pieces which are
cut-constructible and pieces which are not.  One may also calculate
loop amplitudes for massive particles in this way \cite{Massive}, but
cut-free integrals may appear.  The coefficients of these functions
must be determined by other means, such as knowledge of ultraviolet or
infrared divergences.



\section{FACTORIZATION}
\label{FactorizationSection}

In quantum field theory, amplitudes are constrained by their behavior
as kinematic variables vanish; they must factorize into a product of
two amplitudes with an intermediate propagator.  This may be used as a
check on five- or higher-point amplitudes.  (Factorization of
four-point amplitudes in a theory without massive particles
is trivial since the limiting kinematics is degenerate.)
Factorization properties may also be used to help construct new
amplitudes from known ones.  In principle, this can be an extremely
efficient way to obtain amplitudes since one avoids evaluating loop
integrals.

Mangano and Parke have reviewed the factorization properties of
tree-level QCD amplitudes \cite{ManganoReview}.  We shall focus on the
corresponding properties at one loop, which are
a bit more complicated since the amplitudes generally contain infrared
divergent pieces which do not factorize naively.  Nevertheless, as any
kinematic variable vanishes, one-loop amplitudes have a universal
behavior quite similar to that of tree-level amplitudes.

\subsection{\it General Framework}

First we review briefly the situation at tree level.  Color-ordered
amplitudes can have poles only in channels corresponding to the 
sum of {\it cyclicly adjacent} momenta, that is as $P^2_{i,j} \to 0$,
where $P^\mu_{i,j} \equiv (k_i+k_{i+1}+\cdots+k_j)^\mu$.  This is
because singularities arise from propagators going on-shell, and
propagators for color-ordered graphs always carry momenta of the form
$P^\mu_{i,j}$.  The general form of an $n$-point color-ordered tree
amplitude in the limit that $P^2_{1,m}$ vanishes is
$$
\eqalign{
&  A_n^\tree(1,\ldots,n) 
  \mathop{\hbox{\Large$\longrightarrow$}}^{P^2_{1,m} \rightarrow 0}\ \cr
& \hskip .8 cm  
\sum_{\lambda=\pm} A_{m+1}^\tree(1,\ldots,m, P^\lambda)
        {i\over P^2_{1,m} }
        A_{n-m+1}^\tree(m+1,\ldots,n, P^{-\lambda}) \,, \cr}
\equn\label{TreeFactorization}
$$
where $P_{1,m}$ is the intermediate momentum, $A^\tree_{m+1}$ and
$A^\tree_{n-m+1}$ are lower-point scattering amplitudes, and
$\lambda$ denotes the helicity of the intermediate state $P$.  The
intermediate helicity is reversed going from one product amplitude to
the other because of the outgoing-particle helicity convention.

For two-particle channels ($m=2$), \eqn{TreeFactorization} needs to be
modified, because a three-point massless scattering amplitude is not
kinematically possible.  As $P^2_{12}\to0$, $k_1$ and $k_2$ become
collinear.  QCD amplitudes have an angular-momentum obstruction in
this limit.  For example, a gluon of helicity $+1$ cannot split into
two collinear helicity $\pm 1$ gluons and conserve angular momentum.
This transforms the full pole in $P^2_{12} = s_{12}$ into the
square-root of a pole, $1/\sqrt{s_{12}}$, a behavior which is well
captured via the spinor products $\spa1.2,\spb1.2$.  It is useful to
lump all terms not associated with $A_{n-1}^\tree$
in~\eqn{TreeFactorization} into a `splitting amplitude'
$\Split^\tree$.  In particular, as the momenta of adjacent legs $a$
and $b$ become collinear, we have
$$
A_n^\tree(\ldots,a^{\lambda_a},b^{\lambda_b},\ldots)
\ \mathop{\longrightarrow}^{a \parallel b}\
\sum_{\lambda=\pm}
  \Split^\tree_{-\lambda}(z,a^{\lambda_a},b^{\lambda_b}) \,
  A_{n-1}^\tree(\ldots,P^\lambda,\ldots) \,,
\equn\label{TreeSplit}
$$
where $P$ is the intermediate state with momentum $k_P=k_a+k_b$,
$\lambda$ denotes the helicity of $P$, and $z$ is the
longitudinal momentum fraction, $k_a \approx zk_P$, $k_b \approx (1-z)
k_P$.  The universality of these limits can be derived
diagrammatically, but an elegant way to derive it is from string
theory \cite{ManganoReview}, because all the field theory diagrams on
each side of the pole are lumped into one string diagram.  

Given the general form (\ref{TreeSplit}), one may obtain explicit
expressions for the tree-level $g\to gg$ splitting amplitudes from the
four- and five-gluon amplitudes
\cite{TreeCollinear,ManganoReview,TasiLance}.  For example, taking the
collinear limits of~\eqn{gggggmmppptree} for
$A_5^\tree(1^-,2^-,3^+,4^+,5^+)$ and comparing
to~eqs.~(\ref{ggggmmpptree}) and~(\ref{TreeSplit}) shows that
$$
\eqalign{
\Split^\tree_{-}(a^{-},b^{-}) &= 0 \,, \cr
\Split^\tree_{-}(a^{+},b^{+})
            & = {1\over \sqrt{z (1-z)}\spa{a}.b} \,, \cr
\Split^\tree_{+}(a^{+},b^{-})
            & = {(1-z)^2\over \sqrt{z (1-z)}\spa{a}.b} \,, \cr
\Split^\tree_{+}(a^{-},b^{+})
            &= {z^2\over \sqrt{z (1-z)}\spa{a}.b} \,. \cr}
\equn\label{TreeSplittingFunctions}
$$
The remaining helicity configurations are obtained using parity.
The $g\to \qb q$ and $q \to qg$ splitting amplitudes can be 
obtained in similar fashion.

\begin{figure}
\begin{center}
\epsfig{file=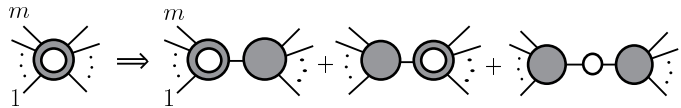,width=3.6in,clip=}
\end{center}
\vskip -.7 cm \caption[]{
\label{MultiFactFigure}  
A schematic representation of the behavior of one-loop amplitudes as
a kinematic invariant vanishes. }
\end{figure}

The situation for color-ordered one-loop amplitudes is similar to
tree level.  The one-loop analog of \eqn{TreeFactorization} is
schematically depicted in \fig{MultiFactFigure}, and is given by
$$
\hskip -.2 cm 
\eqalign{
&  A_n^{\rm loop}(1,\ldots,n) 
  \mathop{\hbox{\Large$\longrightarrow$}}^{P^2_{1,m} \rightarrow 0}\ \cr
& \hskip .2 cm  
\sum_{\lambda=\pm} \biggl[ 
A_{m+1}^{\rm loop} (1,\ldots,m,P^\lambda)
        {i\over P^2_{1,m} }
        A_{n-m+1}^\tree(m+1,\ldots,n,P^{-\lambda}) \cr
& \hskip .4 cm   + 
A_{m+1}^\tree(1,\ldots,m,P^\lambda)
        {i\over P^2_{1,m} }
        A_{n-m+1}^{\rm loop}(m+1,\ldots,n,P^{-\lambda})  \cr
&  \hskip .4 cm  +  
A_{m+1}^\tree(1,\ldots,m,P^\lambda)
        {i \, \Fact_n(1, \!\ldots\!, n)\over P^2_{1,m} }
        A_{n-m+1}^\tree(m+1,\ldots,n,P^{-\lambda})  \biggr]\,, \cr } 
\equn\label{LoopFact}
$$ 
where the one-loop {\it factorization function}, $\Fact_n$, is
independent of helicities and does not cancel the pole in $P^2_{1,m}$.
In an infrared divergent theory, such as QCD, amplitudes do not
factorize `naively': $\Fact_n$ may contain logarithms of kinematic
invariants built out of momenta from {\it both\/} sides of the pole in
$P^2_{1,m}$; $\ln(-s_{n, 1})$ is an example of such a logarithm.  The
factorization functions are nonetheless universal functions depending
on the infrared divergences present in the amplitudes
\cite{Factorization}.

%

\begin{figure}
\begin{center}
\epsfig{file=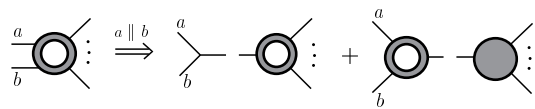,width=2.8in,clip=}
\end{center} 
\vskip -.7 cm \caption[]{
\label{CollinearFigure}  
A schematic representation of the behavior of one-loop amplitudes as
the momenta of two legs become collinear. }
\end{figure}

The collinear limits for color-ordered one-loop amplitudes are a
special case and have the form
$$
\eqalign{
A_{n}^{\rm loop}\ \mathop{\longrightarrow}^{a \parallel b}\
\sum_{\lambda=\pm}  \biggl(
 & \Split^\tree_{-\lambda}   (z, a^{\lambda_a},b^{\lambda_b})\,
      A_{n-1}^{\rm loop}(\ldots(a+b)^\lambda\ldots) \cr
& + \Split^{\rm loop}_{-\lambda}(z,a^{\lambda_a},b^{\lambda_b})\,
      A_{n-1}^\tree(\ldots(a+b)^\lambda\ldots) \biggr) \,,
\cr}
\equn\label{LoopSplit}
$$
which is schematically depicted in \fig{CollinearFigure}.  The
splitting amplitudes
$\Split^\tree_{-\lambda}(a^{\lambda_a},b^{\lambda_b})$ and
$\Split^{\rm loop}_{-\lambda}(a^{\lambda_a},b^{\lambda_b})$ are
universal: they depend only on the two momenta becoming collinear, and
not upon the specific amplitude under consideration.  The explicit
$\Split^{\rm loop}_{-\lambda}(a^{\lambda_a},b^{\lambda_b})$ were
originally determined from the four- and five-point one-loop
amplitudes \cite{FiveGluon,Fermion} in much the same way as we
obtained the tree-level splitting amplitudes above.  (See appendix~B
of ref.~\cite{SusyFour}.)  Soft limits --- the behavior as any
particular $k_i \rightarrow 0$ --- are also useful for constraining
the form of one-loop amplitudes, and have a form analogous to
\eqn{LoopSplit}.

%

In performing explicit calculations, factorization provides an
extremely stringent check since one must obtain the correct limits in
all channels.  A sign or labeling error, for example, will invariably
be detected in some limits.  In some cases one can also use
factorization to construct ans\"atze for higher-point amplitudes
\cite{AllPlus,SusyFour}.  One writes down a sufficiently general form
for a higher-point amplitude, containing arbitrary coefficients which
are then fixed by imposing the correct behavior as kinematic variables
vanish.  A collinear bootstrap of this form would, however, miss
functions that are nonsingular in all collinear limits.  For
five-point amplitudes it is possible to write down such a function,
namely
$$
{\varepsilon(1,2,3,4)
\over \spa1.2 \spa2.3 \spa3.4 \spa4.5 \spa5.1}
\, ,
\equn\label{FivePointAmbiguity}
$$
since the contracted antisymmetric tensor $\varepsilon(1,2,3,4) \equiv
4i \varepsilon_{\mu\nu\rho\sigma} k_1^\mu k_2^\nu k_3^\rho k_4^\sigma$
vanishes when any two of the five vectors $k_i$ become collinear
(using $\sum_{i=1}^5 k_i = 0$).  However, it is quite possible that the
factorization constraint uniquely specifies the rational functions of
color-ordered $n\ge 6$-point amplitudes, given the lower point
amplitudes. A heuristic explanation of this conjecture is that as the
number of external legs increases, by dimensional analysis the
amplitudes require ever increasing powers of momenta in the
denominators.  Thus one expects more kinematic poles from the
denominator than zeros from the numerator.  We know of no counter-examples to
this conjecture, but don't have a proof either.

\subsection{\it Examples}

As an example of the behavior of a one-loop amplitude in
a collinear limit, consider the $N=4$ five-gluon amplitude
$A_{5;1}^\neqfour$ given in~\eqn{gggggmmppploop}. 
Taking the limit $k_4 \rightarrow zP, \, k_5 \rightarrow (1-z)P$,
and using the four-gluon result~(\ref{ggggmmppneqfour}), we find
$$
\eqalign{
A_{5;1}^\neqfour(1^-,2^-,3^+,4^+,5^+)  
\mathop{\longrightarrow}^{4 \parallel 5}\
& \Split^\tree_-(4^+, 5^+) A_{4;1}^\neqfour(1^-, 2^-, 3^+, P^+) \cr
& + \Split^\neqfour_-(4^+, 5^+) A_{4}^\tree (1^-, 2^-, 3^+, P^+) \,,\cr}
\equn
$$
where the tree splitting amplitude is given in
\eqn{TreeSplittingFunctions}. This limit determines the one-loop $N=4$
multiplet contribution to the $g\to gg$ splitting amplitude,
$$
\eqalign{
&\Split^\neqfour_-(a^+, b^+)  = \cr
& \hskip 7 mm 
\cg \, \Split^\tree_-(a^+, b^+) 
\biggl[ - {1 \over \eps^2} (-s_{ab} z (1-z))^{-\eps}
  \!+\! 2 \ln z \ln(1-z) 
  \!-\! {\pi^2 \over 6} \biggr] \,. \cr}
\equn
$$
Note that the loop splitting amplitude has
absorbed the mismatch of infrared divergences between the five- and
four-point amplitudes.  This splitting amplitude will reappear in
different collinear limits of this and other amplitudes.

Given the splitting amplitudes and five-point amplitudes, it is
also possible to construct conjectures for higher-point amplitudes by
demanding that they factorize correctly.  Consider, for example, the
complex-scalar loop contribution to a five-gluon amplitude with all
identical helicities \cite{FiveGluon},
$$
\eqalign{
A_{5;1}^{\rm scalar} & (1^+,2^+,3^+,4^+,5^+) \cr
& = {i\over 96\pi^2}\,
  {  s_{12}s_{23} + s_{23}s_{34} + s_{34}s_{45} + s_{45}s_{51} +
     s_{51}s_{12} + \pol(1,2,3,4)
   \over \spa1.2 \spa2.3 \spa3.4 \spa4.5 \spa5.1 }\, .  \cr }
\equn\label{AllPlusFive}
$$
As noted in section~\ref{SusyDecompositionSubsection}, for this
helicity configuration the gluon and fermion loops are proportional to
the scalar-loop contribution.  One can verify that this amplitude has
the correct collinear limits (\ref{LoopSplit}), using the four-gluon
amplitude (\ref{FourPlusResult}).

Using \eqn{LoopSplit}, the explicit form of the tree splitting
amplitudes (\ref{TreeSplittingFunctions}), $A^{\rm tree}_n (1^\pm,
2^+, \cdots, n^+) = 0$, and experimenting at small $n$, we can
construct higher-point amplitudes by writing down general forms with
only two-particle poles, and requiring that they have the correct
collinear limits.  Doing so leads to the all-$n$ ansatz
\cite{AllPlus},
$$
\eqalign{
A_{n;1}^{\rm scalar}(1^+,2^+,\ldots,n^+)\ =\ -{i \over 48\pi^2}\,
\sum_{1\leq i_1 < i_2 < i_3 < i_4 \leq n}
{ {\rm tr}_-[i_1 i_2 i_3 i_4]
\over \spa1.2 \spa2.3 \cdots \spa{n}.1 } \,,}
\equn
$$
where $\tr_-[i_1 i_2 i_3 i_4] = {1\over 2}\tr_-[(1-\gamma_5)
\ksl_{i_1} \ksl_{i_2} \ksl_{i_3} \ksl_{i_4}]$.  This result has 
been confirmed by Mahlon via recursive techniques~\cite{Mahlon}.

Indeed, the infinite sequence of one-loop $N=4$ supersymmetric MHV
amplitudes was first constructed via a collinear bootstrap and only
then calculated using the unitarity method described in
section~\ref{SusyExamplesSubsection}. Other helicity configurations
are more complicated, due to the appearance of multi-particle poles.
Nevertheless, one can construct some six-point amplitudes from
knowledge of the five-point amplitudes.  This is most useful for the
rational-function parts, which can be obtained via unitarity only by
working to higher order in~$\eps$.



\section{CONCLUSIONS AND OUTLOOK}

We have reviewed various developments in calculational techniques for
one-loop gauge theory amplitudes, especially in QCD.  Such
calculations are necessary in order to confront theoretical
predictions with experiments to some degree of precision.  Feynman
rules, however, become extremely cumbersome for one-loop
multi-parton calculations. 
Even the simplest processes are rather difficult to
calculate without aid of a computer and for five or more external legs
traditional methods break down because of an exponential explosion in
algebra.  The results, however, are usually quite compact, especially
when compared to intermediate expressions.

The computational situation can be greatly improved by combining a
number of ideas.  Methods that have previously been used at
tree-level, such as spinor helicity \cite{SpinorHelicity}, color
decomposition \cite{Color}, and supersymmetry Ward identities
\cite{OldSWI,NewSWI}, remain very useful at one loop. String theory
motivates a number of improved organizational ideas such as
supersymmetric decompositions, relations between color-decomposed
amplitudes and improved gauge choices
\cite{Long,Mapping,StringBased,TasiZvi}.  These ideas mesh nicely with
the use of Cutkosky rules \cite{Cutting} to obtain complete
amplitudes.  In the superstring organization of the amplitude,
components can be identified whose rational as well as cut-containing
parts can be obtained directly from knowledge of the branch cuts
\cite{SusyFour,SusyOne}.  The remaining components, though more
difficult, can be attacked either by evaluating cuts to higher order
in $\eps$ or by exploiting universal factorization properties 
\cite{AllPlus}.

The techniques discussed in this review have made possible a variety
of new calculations, including those of all five-parton amplitudes
\cite{FiveGluon,Zoltanqqqqg,Fermion} and of certain infinite sequences
of massless amplitudes.  The methods have also been applied to
amplitudes containing massive particles \cite{WeakInt,Massive} and to
gravitational amplitudes \cite{Gravity}.  Mahlon has also used
recursion relations, outside of the scope of this review, to obtain
infinite sequences of fermion loop amplitudes with maximal helicity
violation \cite{Mahlon}.

It would be desirable to extend these techniques to
two-loop multi-parton calculations; while various authors have taken
first steps \cite{TwoLoopUnitarity,TwoLoopQED,TwoLoopStrings,%
TwoLoopIntegrals} in this direction, a great deal of work remains to
be done.

\vskip .25 cm 

We thank G. Chalmers, A.G.~Morgan and especially D.C.~Dunbar for
collaboration on work described in this review.  This work was
supported in part by the US Department of Energy under grants
DE-FG03-91ER40662 and DE-AC03-76SF00515, by the Alfred P. Sloan
Foundation under grant BR-3222, and by the {\it Direction des Sciences
de la Mati\`ere\/} of the {\it Commissariat \`a l'Energie Atomique\/}
of France.

\vfill
\eject
\bigskip


\end{document}